\documentclass[12pt,preprint,english]{aastex}
\usepackage[T1]{fontenc}
\usepackage{amssymb}
\makeatletter
\slugcomment{}
\shorttitle{}
\shortauthors{}
\usepackage{natbib}
\makeatother
\received{2003 September 16}
\begin{document}

\title{Tracing the Magnetic Field in Orion A}

\author{Martin Houde\altaffilmark{1}}

\email{houde@submm.caltech.edu}

\author{C. Darren Dowell\altaffilmark{2,3}, Roger H. Hildebrand\altaffilmark{4,5},
Jessie L. Dotson\altaffilmark{6}, John E. Vaillancourt\altaffilmark{7},
Thomas G. Phillips\altaffilmark{3}, Ruisheng Peng\altaffilmark{1},
Pierre Bastien\altaffilmark{8}}

\altaffiltext{1}{Caltech Submillimeter Observatory, 111 Nowelo Street, Hilo, HI 96720}

\altaffiltext{2}{Jet Propulsion Laboratory, 4800 Oak Grove Drive, Pasadena, CA 91109}

\altaffiltext{3}{California Institute of Technology, Pasadena, CA 91125}

\altaffiltext{4}{Department of Astronomy and Astrophysics and Enrico Fermi Institute, University of Chicago, Chicago, IL 60637}

\altaffiltext{5}{Department of Physics, University of Chicago, Chicago, IL 60637}

\altaffiltext{6}{NASA Ames Research Center, Moffet Field, CA 94035}

\altaffiltext{7}{Physics Department, University of Wisconsin, 1150 University Avenue, Madison, WI 53706}

\altaffiltext{8}{Département de Physique, Université de Montréal, Montréal, Québec H3C 3J7, Canada} 

\begin{abstract}
We use extensive 350 $\mu$m polarimetry and continuum maps obtained
with Hertz and SHARC II along with HCN and HCO$^{+}$ spectroscopic
data to trace the orientation of the magnetic field in the Orion A
star-forming region. Using the polarimetry data, we find that the
direction of the projection of the magnetic field in the plane of
the sky relative to the orientation of the integral-shaped filament
varies considerably as one moves from north to south. While in IRAS
05327-0457 and OMC-3 MMS 1-6 the projection of the field is primarily
perpendicular to the filament it becomes better aligned with it at
OMC-3 MMS 8-9 and well aligned with it at OMC-2 FIR 6. The OMC-2 FIR
4 cloud, located between the last two, is a peculiar object where
we find almost no polarization. There is a relatively sharp boundary
within its core where two adjacent regions exhibiting differing polarization
angles merge. The projected angle of the field is more complicated
in OMC-1 where it exhibits smooth variations in its orientation across
the face of this massive complex. We also note that while the relative
orientation of the projected angle of the magnetic field to the filament
varies significantly in the OMC-3 and OMC-2 regions, its orientation
relative to a fixed position on the sky shows much more stability.
This suggests that, perhaps, the orientation of the field is relatively
unaffected by the mass condensations present in these parts of the
molecular cloud. By combining the polarimetry and spectroscopic data
we were able to measure a set of average values for the inclination
angle of the magnetic field relative to the line of sight. We find
that the field is oriented quite close to the plane of the sky in
most places. More precisely, the inclination of the magnetic field
is $\approx73^{\circ}$ around OMC-3 MMS 6, $\approx74^{\circ}$ at
OMC-3 MMS 8-9, $\approx80^{\circ}$ at OMC-2 FIR 4, $\approx65^{\circ}$
in the northeastern part of OMC-1, and $\approx49^{\circ}$ in the
Bar. The small difference in the inclination of the field between
OMC-3 and OMC-2 seems to strengthen the idea that the orientation
of the magnetic field is relatively unaffected by the agglomeration
of matter located in these regions. We also present polarimetry data
for the OMC-4 region located some $13\arcmin$ south of OMC-1.
\end{abstract}


\section{Introduction}

The determination of the orientation of the magnetic field in molecular
clouds and their surroundings is important for the assessment of its role
and relative importance in the evolution of the clouds and in the
process of star formation. Numerous scenarios have been proposed concerning
the different stages involved in the birth of stars from primordial
clouds \citep{Shu 1987, McKee 1993, Heiles 1993, Mouschovias 1999, Williams 2000}.
But, whatever the details, they all seek to describe at some level
or another the interaction of the magnetic field with its environment
and its influence during gravitational collapse. The magnetic field
will resist and slow down the collapse and perhaps even shape the
structures observed within the clouds themselves \citep{Fiege 2000,Matthews 2001}.
Conversely, measurements of the orientation of the magnetic field
in molecular clouds and prestellar cores can tell us something about
the ways in which cores form; e.g., an hourglass geometry is usually
interpreted as a sign of the magnetic field having been dragged by
the gas during its collapse (see for example \citet{Schleuning 1998}).
In that respect, data obtained from polarimetry measurements at far-infrared
and submillimeter wavelengths which trace the orientation of the projection
of the magnetic field in the plane of the sky \citep{Hildebrand 1988}
are essential for this task and for creating, testing and refining
models \citep{Matthews 2001}.

Recently, \citet{Houde 2002} have proposed a new technique that would
allow for a more complete determination of the orientation of the
magnetic field in that it renders possible the measurement of the
inclination or viewing angle that the field makes relative to the
line of sight. This is accomplished by combining polarimetry data
and ion-to-neutral line width ratios obtained from the spectra of
coexistent molecular species. The technique was first applied to the
M17 molecular cloud using an extensive 350 $\mu$m Hertz polarimetry
map and HCO$^{+}$/HCN spectroscopic data \citep{Houde 2000a, Houde 2000b, Houde 2002},
all obtained at the Caltech Submillimeter Observatory (CSO). 

We pursue a similar task in this paper where our goal is primarily
to use polarimetry and spectroscopic data (as well as continuum maps)
to trace the orientation of the magnetic field along the integral-shaped
filament (ISF; see \citet{Bally 1987}) of the Orion A molecular cloud
(OMC). We will use the 350 $\mu$m polarimetry maps obtained with
the Hertz polarimeter \citep{Dowell 1998}, superposed on SHARC II
continuum maps at the same wavelength \citep{Dowell 2003}, to study
the orientation of the projection of the magnetic field in the plane
of the sky along with HCO$^{+}$ and HCN spectra to evaluate the inclination
of the field relative to the line of sight.

\setcounter{footnote}{0}

\section{Observations}

The core of the data presented in this paper comes from extensive
350 $\mu$m polarimetry maps obtained with the Hertz polarimeter \citep{Dowell 1998}
during numerous nights of observations at the Caltech Submillimeter
Observatory. In all, seven different regions of the Orion A molecular
cloud were studied. They are from north to south: \emph{i)} IRAS 05327-0457,
observed on 2002 February 17-18, \emph{ii)} OMC-3 MMS 1-6, observed
on 1997 September 19, 1998 February 15, 1999 January 27 and 2002 February
18, \emph{iii)} OMC-3 MMS 8-9, observed on 2002 February 18, \emph{iv)}
the neighborhood of OMC-2 FIR 3-4, observed on 1997 September 19,
1998 February 15 and 1999 January 27, \emph{v)} the neighborhood of
OMC-2 FIR 6, observed on 2002 February 18, \emph{vi)} OMC-1, observed
on numerous nights from 1998 February to 2002 February, and finally
\emph{vii)} OMC-4, observed on 2001 April 16 and 2002 February 18.
Although Hertz data provide both polarimetry and continuum maps, we
use a SHARC II map for illustrations of the 350 $\mu$m continuum
\citep{Dowell 2003}. SHARC II has a better spatial resolution than
Hertz ($\approx12\arcsec$ compared to $\approx20\arcsec$) and the
data obtained with it and presented here continuously cover all of
the regions mentioned above, unlike the Hertz maps. The SHARC II data
were obtained on three separate nights: 2002 November 24, 2003 March
1 and 2003 April 16.

The Hertz data were acquired with conventional azimuth chopping and
nodding. The chopper throw ranged from $5\arcmin$ to $8.4\arcmin$,
and we made attempts to avoid chopping into known emission sources
by choosing the hour angle of the observations. The SHARC II data
were acquired with crossing linear scans without chopping the secondary
mirror. The data analysis technique subtracts from the source emission
a model for the atmospheric emission which is improved through iterations.
The total integration time was 3.7 hours in conditions with moderately
low zenith opacity ($\tau_{350\mu{\rm {m}}}\sim1.5$).

We present in Figure \ref{cap:omc3} the northern portion of the data
where the polarimetry measurements for IRAS 05327-0457 and OMC-3 are
plotted on top of the SHARC II continuum map. Figures \ref{cap:omc2},
\ref{cap:omc1} and \ref{cap:omc4} present the same type of data
but for the OMC-2, OMC-1 and OMC-4 regions, respectively. For each
of these figures, the thick (thin) polarization vectors have a polarization
level and uncertainty such that $P\geq3\sigma_{P}$ ($P\geq2\sigma_{P}$),
circles indicate cases where $P+2\sigma_{P}<1\%$ and $P<2\sigma$,
and the darker polarization vectors and circles denote positions where
spectroscopic data were also obtained. The beam widths are shown in
the lower left corner, with the solid and open circles for SHARC II
and Hertz, respectively. The complete unbroken SHARC II continuum
map can be seen in Figure \ref{cap:oriona}. Because of the very low
levels of polarization observed on some of the sources studied in
this paper, we have decided to include data points that have an uncertainty
in their polarization level better or equal to $2\sigma$ instead
of the usual $3\sigma$ common to Hertz papers (see for example \citet{Houde 2002}).
The set of polarization angles measured for these $2\sigma$ points
have an uncertainty $\la15^{\circ}$, as opposed to $\la10^{\circ}$
for $3\sigma$ points, and agree well with adjacent points of lower
uncertainty. Their inclusion will not affect the conclusions drawn
later. Details of the polarimetry data will be found in Tables \ref{ta:omc1}
to \ref{ta:omc4}.

We have also measured HCN and HCO$^{+}$ spectra in the $J=4\rightarrow3$
transition at different positions along the ISF; these are necessary
to evaluate the inclination angle of the magnetic field relative to
the line of sight. In all, twenty-seven such pairs of spectra were
obtained using the 300-400 GHz Receiver of the CSO: three in the Bar,
seven in the northeastern part of OMC-1, seven in OMC-2 FIR 4, five
in and around OMC-3 MMS 6 and five in OMC-3 MMS 8-9. In the latter,
two of the five points do not have corresponding polarimetry vectors
with $P\geq2\sigma_{P}$ or $P+2\sigma_{P}<1\%$ and are identified
with dark dots in Figure \ref{cap:omc3}. Typical examples of measured
HCN and HCO$^{+}$ spectra, along with a fit to their profiles, can
be found in Figure \ref{cap:spectra}. This set of data was obtained
during numerous nights of observation at the CSO on 2002 September,
2002 December, 2003 January to 2003 April and 2003 August.

\section{Results}

\subsection{The 350 $\mu$m SHARC II map}

Although our analysis will  concentrate on the nature of the magnetic
field in Orion A, it is appropriate at this point to say a few words
concerning the 350 $\mu$m SHARC II continuum map presented in Figure
\ref{cap:oriona}. The Orion A molecular cloud has been mapped at
several wavelengths in the past and we do not intend to repeat the
work presented in other papers. For example, \citet{Lis 1998} also
published a 350 $\mu$m map of this region, obtained with SHARC, and
gave a detailed analysis of the physical conditions found in the different
parts of the cloud as well as a list (in their Table 1) of the sources
that can be found in OMC-2 and OMC-3 (see also \citet{Chini 1997}
for observations at 1.3 mm of the OMC-2 and OMC-3 regions). Our map
covers the same area as that of \citet{Lis 1998} but also extends
further both north of OMC-3 and south of OMC-1, and is wider in its
east-west coverage. More precisely, our map includes the neighborhood
of IRAS 05327-0457 \citep{Makinen 1985, Mookerjea 2000a, Mookerjea 2000b}
which extends some $\approx4\arcmin$ north of OMC-3 MMS 1. It is
in fact seen to be an extension of the OMC-3 MMS 1-6 filament that
branches off to the north in a region of higher flux located at $\Delta\alpha\simeq\mathrm{-3}\arcmin$
and $\Delta\delta\simeq\mathrm{1}7\arcmin$ on Figure \ref{cap:oriona}.
This extension then turns southeast in a direction parallel to the
OMC-3 MMS 1-6 complex and ends up at the CSO 1 condensation of \citet{Lis 1998}
at $\Delta\alpha\simeq\mathrm{3}\arcmin$ and $\Delta\delta\simeq\mathrm{14.4}\arcmin$
(better seen on Figure \ref{cap:omc3}). This IRAS 05327-0457 region
was also part of an extensive 850 $\mu$m SCUBA map published by \citet{Johnstone 1999}
and our results are seen to be in qualitative agreement with theirs.
The same can be said of the V-shaped OMC-4 region \citep{Johnstone 1999},
located some $\approx13\arcmin$ south of the KL Nebula of OMC-1,
and of the whole extent of the two maps, i.e., theirs and ours, which
have a similar coverage.

The sensitivity varies somewhat across the SHARC II image due to variable
integration time and weather conditions. For most of the regions including
OMC-1, OMC-2 and OMC-3, the RMS is approximately $1\,{\rm {Jy}}/12\arcsec$
beam, with higher noise at the edges of the rectangular region of
coverage. For the IRAS05327-0457 and OMC-4 fields, the RMS is approximately
$0.3\,{\rm {Jy}}/12\arcsec$ beam.

\subsection{The polarimetry data}

\subsubsection{The IRAS 05327-0457 and OMC-3 regions}

Figure \ref{cap:omc3} shows a close-up of the IRAS 05325-0457 and
OMC-3 fields from the SHARC II map along with the Hertz polarization
vectors obtained in three different regions: IRAS 05327-0457, OMC-3
MMS 1-6 and OMC-3 MMS 8-9. One thing to notice about the first two
regions is that the general orientation of the polarization vectors
 is in both cases well aligned with, or similarly that the magnetic
field is practically perpendicular to the local filaments. Indeed,
if we visually estimate the orientation of the filaments, we find
that both IRAS 05327-0457 and OMC-3 MMS 1-6 are aligned at approximately
$132^{\circ}$ (east from north, as will always be the case for all
angles measured in the plane of the sky). On the other hand, a simple
arithmetic mean of the polarization angle ($PA$) of the vectors in
both regions gives $\simeq142^{\circ}\pm18^{\circ}$ for IRAS 05327-0457
and $\simeq137^{\circ}\pm9^{\circ}$ for OMC-3 MMS 1-6, respectively.
These values are practically unchanged if we instead average the Stokes
parameters from the same ensembles of points, we then get $\left\langle PA\right\rangle \simeq143^{\circ}\pm3^{\circ}$
for IRAS 05327-0457 and $\left\langle PA\right\rangle \simeq136^{\circ}\pm1^{\circ}$
for OMC-3 MMS 1-6, respectively. Since these two types of averages
usually agree well, and that we will later deal with objects exhibiting
low polarization levels, we will for now on only use Stokes averages.
The difference between the orientation of the polarization vectors
and the local filaments is on average only approximately $5^{\circ}$
to $10^{\circ}$ for these two regions, i.e., they are well aligned
with each other. A similar result has already been reported for the
OMC-3 MMS 1-6 region by \citet{Matthews 2001} using SCUBA at 850
$\mu$m (see their Figure 2). The fact that the same is true for IRAS
05327-0457 adds credence to their suggestion that perhaps the ordered
structure of the magnetic field has not been disturbed or deflected
by the gravitational collapses of the protostellar cores present in
these regions \citep{Chini 1997}.

OMC-3 MMS 8-9 is the third region in Figure \ref{cap:omc3} where
we have obtained polarimetry data. As was noticed by \citet{Matthews 2001}
at 850 $\mu$m, we find that, as compared to the case of OMC-3 MMS
1-6, the orientation of the polarization vectors has significantly
changed in relation to the local filament. This part of the ISF is
oriented at $\approx8^{\circ}$, roughly a $60^{\circ}$ shift relative
to OMC-3 MMS 1-6. The polarization angle calculated with the average
of the Stokes parameters yields $\left\langle PA\right\rangle \simeq122^{\circ}\pm2^{\circ}$.
But since OMC-3 MMS 8-9 exhibits a significant amount of flux both
east and west of the main core, a case could be made in favor of limiting
ourselves only to vectors located strictly in the core (i.e., the
polarization vectors situated on the ten positions of highest local
flux around $\Delta\alpha\simeq2\arcmin$ and $\Delta\delta\simeq7.5\arcmin$)
before calculating the Stokes averages. It turns out, however, that
the change is not significant when we do so as we then obtain $\left\langle PA\right\rangle \simeq129^{\circ}\pm3^{\circ}$.
Therefore, depending on which number we use we find that the vectors
are misaligned by approximately $114^{\circ}$ to $121^{\circ}$ in
relation to the filament. This is somewhat more than the mean of $86^{\circ}$
measured by \citet{Matthews 2001} at 850 $\mu$m; the difference
may lie in the fact that our vectors located east of the main core
do not change appreciably from those located in the core, as seems
to be the case for the SCUBA data. At any rate, this represents a
significant change when compared to what was obtained in IRAS 05327-0457
and OMC-3 MMS 1-6.

But perhaps the most important thing to stress from these results
is that the average polarization angle does not change much from IRAS
05327-0457 to OMC-3 MMS 1 to OMC-3 MMS 8- 9. Indeed the change is
only $\approx13^{\circ}$. Since it is believed that, at the wavelengths
dealt with here, the orientation of the magnetic field projected in
the plane of the sky is perpendicular to that of the polarization
vectors \citep{Hildebrand 1988}, the same can be said of the projected
direction of the field. One must somewhat temper this assertion as
there is some indication from the 850 $\mu$m data of \citet{Matthews 2001}
that the orientation of the polarization vectors measured between
OMC-3 MMS 6 and OMC-3 MMS 7 follows the changing orientation of the
filament. This  could increase the dispersion in the orientation of
the polarization vectors when measured over the field presented in
Figure \ref{cap:omc3}. 

Finally, we end this section with a few words concerning the polarization
levels detected in these parts of the ISF. We find that the percentage
of polarization is typical of what is usually observed with Hertz
at 350 $\mu$m as $0\%\lesssim P\lesssim5\%$, with the highest levels
detected in IRAS 05327-0457 and the lowest levels in OMC-3 MMS 8-9.
This is also reflected in the Stokes averaged polarization levels,
which are $2.33\%\pm0.37\%$, $1.55\%\pm0.12\%$ and $1.02\%\pm0.10\%$
for IRAS 05327-0457, OMC-3 MMS 1-6 and OMC-3 MMS 8-9, respectively.

\subsubsection{The OMC-2 region}

We present with Figure \ref{cap:omc2} the section of the SHARC II
map that is centered on the OMC-2 region along with the polarization
data obtained with Hertz in and around OMC-2 FIR 3-4 and FIR 6. This
region is characterized, amongst other things, by a gradual change
in the orientation of the filament. While in the northern part of
the map the filament makes an angle of $\approx160^{\circ}$ relative
to north, it goes through the north-south orientation at OMC-2 FIR
4 and ends at $\approx35^{\circ}$ (or $\approx215^{\circ}$) at OMC-2
FIR 6. A visual inspection of the map allows us to quickly evaluate
the general characteristics of the polarimetry data. First, just north
of OMC-2 FIR 4 the polarization vectors seem to be well aligned with
the filament but not with the vectors measured in the OMC-3 field.
Second, OMC-2 FIR 4 displays very little polarization which at first
renders it impossible to safely determine any relative alignment in
this region. This is probably the object which exhibits the lowest
levels of polarization amongst all of those observed with Hertz so
far. Although a strong depolarization effect, defined by a systematic
decrease in polarization level with increasing continuum flux, is
commonly seen in this type of study \citep{Dotson 1996, Matthews 2001, Houde 2002},
this is probably the most severe case to date. Finally, the orientation
of the vectors changes one more time around OMC-2 FIR 6 where they
appear to be perpendicular to the filament.

We can quantify these observations by, once again, calculating the
averaged Stokes parameters (it would be unwise to attempt anything
else with the OMC-2 FIR 4 data) and deriving values for the polarization
angle in these parts of the ISF. We find that $\left\langle PA\right\rangle \simeq175^{\circ}\pm3^{\circ}$
for the region adjacent to and north of OMC-2 FIR 3 (for $\Delta\delta\mathrm{>3.4}\arcmin$),
$\left\langle PA\right\rangle \simeq121^{\circ}\pm5^{\circ}$ for
the core of OMC-2 FIR 4 (for $2.8\arcmin<\Delta\delta\mathrm{<3.4}\arcmin$)
and $\left\langle PA\right\rangle \simeq115^{\circ}\pm6^{\circ}$
for OMC-2 FIR 6. These give the following differences between the
mean orientation of the vectors and the filament: $\approx15^{\circ}$,
$\approx59^{\circ}$ and $\approx80^{\circ}$, in the same order. 

It is interesting to note that there appears to be a relatively sharp
transition close to the position of peak intensity of OMC-2 FIR 4
($\Delta\delta\simeq\mathrm{3.}1\arcmin$) where the polarization
angle changes abruptly from its FIR 6 value ($\approx115^{\circ}$,
i.e., almost perpendicular to the local orientation of the filament)
to the FIR 3 value ($\approx175^{\circ}$, i.e., almost parallel to
the local orientation of the filament). One might hope to explain
the unusually low level of polarization measured at OMC-2 FIR 4 by
combining the measurements obtained on either side of this boundary.
We could expect, for example, a drop in the polarization level at
OMC-2 FIR 4 if the polarization patterns north and south of it preserve
their relative orientation to the filament as they merge toward it
(for comparable polarized flux). The Stokes averaged polarization
levels calculated for OMC-2 FIR 6 and OMC-2 FIR 3 are $0.91\%\pm0.28\%$
and $0.73\%\pm0.14\%$, respectively; we were unable to reproduce
the result obtained for FIR 4 ($0.35\%\pm0.08\%$) by combining them
(along with their corresponding mean polarization angle and flux).

As was the case for the OMC-3 region, we find that the orientation
of the average polarization angle does not vary much between OMC-2
FIR 4 and OMC-2 FIR 6 ($\approx6^{\circ}$) and is also not very different
to the values measured in OMC-3. In fact, neglecting the field adjacent
to and north of OMC-2 FIR 3, we find a smooth decrease from the north
of OMC-3 (or IRAS 05327-0457) to the south of OMC-2. 

We note finally that the boundary where the jump in the orientation
of the polarization angle occurs is almost coincident with a region
in the vicinity of OMC-2 FIR 3 where intense outflow activity is detected.
According to \citet{Williams 2003}, this source is composed of a
binary which drives a pair of criss-crossed flows oriented at $\approx30^{\circ}$,
similar to the orientation of the projection of the magnetic field
in the plane of the sky at and south of OMC-2 FIR 4. Moreover, there
is also some evidence for another flow, in the same neighborhood,
oriented at $\approx80^{\circ}$ in the plane of the sky (J. P. Williams,
private communication), this is again very close to the orientation
of the field at and north of OMC-2 FIR 3.

\subsubsection{The OMC-1 and the Bar regions}

We present in Figure \ref{cap:omc1} a section of the SHARC II map
centered on the OMC-1 region along with the polarization data obtained
with Hertz superposed on the continuum. A quick glance will suffice
to convince the reader of the totally different orientation of the
polarization vectors measured in this field when compared to those
of the OMC-3 and OMC-2 regions. A small subset of the polarimetry
data, centered in the neighborhood of the KL Nebula (at $\Delta\alpha\simeq\mathrm{-0.7}\arcmin$
and $\Delta\delta\simeq\mathrm{-9.4}\arcmin$ on Figure \ref{cap:omc1}),
has already been published by \citet{Schleuning 1998} along with
an extensive 100 $\mu$m map, albeit at a lower resolution, of the
OMC-1 cloud. The polarization pattern was then interpreted as being
consistent with that produced by a magnetic field shaped like an hourglass
(see Figure 4 of \citet{Schleuning 1998}) resulting from the gravitational
distortion caused by the IRc 2 Ridge. Eight positions in OMC-1 have
also been measured in polarimetry by \citet{Vallée 1999} at 760 $\mu$m.
The polarization angles are quite comparable, within uncertainties,
except for two positions which do not agree. Usually, the polarization
levels are also similar, or somewhat larger at 760 $\mu$m (within
a beam width of $\approx14\arcsec$).

Although our 350 $\mu$m polarimetry map qualitatively agrees well
with the 100 $\mu$m counterpart, there are a few differences and
features in the new set of data that complicate the interpretation.
This is especially true in the southern part of the map, most notably
in and around the Bar where there is a significant change in the orientation
of the vectors (more on this below) that does not fit with the hourglass
interpretation. There is also a difference in the region east of the
Ridge where the magnetic field on our map appears to be basically
aligned along an east-west axis. But, all in all, there is an unmistakable
pinch in the orientation of the magnetic field in the neighborhood
of the Ridge with the region of largest curvature located just north
of the KL Nebula. But as was stated by \citet{Rao 1998}, caution
must be used in interpreting the orientation of the magnetic field
from polarimetry data in regions of high outflow activity, as is the
case for the Ridge. This is owed to the possibility that, in the presence
of outflows, grains can be aligned with their long axis parallel to
the magnetic field through the Gold alignment mechanism \citep{Gold 1952, Lazarian 1994, Lazarian 1997}.
Incidentally, our data also show the existence of the {}``polarization
hole'' in the vicinity of the KL Nebula that was shown by \citet{Rao 1998}
to be caused by an abrupt and small scale $90^{\circ}$ change in
the orientation of the polarization vectors and presumably caused
by the presence of outflows, as was just mentioned. This, however,
is not the only place where a local decrease of the polarization level
is seen. The same is true, for example, in the neighborhood of \emph{i)}
the Bar, \emph{ii)} in a relatively large region in the south of the
map, and west of the Bar, centered at $\Delta\alpha\simeq\mathrm{-1.0}\arcmin$
and $\Delta\delta\simeq\mathrm{-12.4}\arcmin$ or again \emph{iii)}
some $2\arcmin$ north of KL at $\Delta\alpha\simeq\mathrm{-0.2}\arcmin$
and $\Delta\delta\simeq\mathrm{-7.5}\arcmin$ (a local peak in the
continuum flux).

As was said earlier, the overall orientation of the polarization vectors
in OMC-1 is totally different from what we have seen so far in Orion
A. By doing averages of Stokes parameters on small ensembles of points
we find that the polarization angle changes significantly as one moves
about the cloud. Although in the north of the map the vectors converge
to an orientation of $\approx40^{\circ}$ (somewhat different from
the $\approx30^{\circ}$ measured around the KL Nebula, in agreement
with previous results (see Table 1 of \citet{Schleuning 1998}), the
polarization angle is seen to cover values anywhere from $\approx0^{\circ}$
east of the Ridge to $\approx60^{\circ}$ in the northwestern corner
of the map to $\approx160^{\circ}$ in the southwestern corner. The
southern part of the map, and the Bar in particular, is where we find
the most abrupt changes in the orientation of the vectors. A Stokes
average of eighteen data points located in the Bar gives a mean polarization
angle of $\left\langle PA\right\rangle \simeq82^{\circ}\pm4^{\circ}$.

The polarization levels also greatly vary across OMC-1. The lowest
levels are found in the south of the map where a few positions with
no polarization (i.e., where $P+2\sigma_{P}<1\%$ and $P<2\sigma_{P}$)
are found. The mean polarization level found in the Bar, using the
same eighteen points as before, is $\left\langle P\right\rangle =0.76\%\pm0.14\%$.
\citet{Schleuning 1998} interpreted this low level as being caused
by a local orientation of the magnetic field along the line of sight
as a result of being pushed out of the M42 H {\small II} region. But
as will be discussed in section \ref{sub:Calibration}, our measurements
for the inclination of the magnetic field in this region are not in
agreement with this interpretation. An alternative explanation could
reside in the fact that the dust is colder in the Bar than in other
regions of higher flux in OMC-1, thus its lower levels of polarization
would be consistent with the hypothesis that cooler dust is intrinsically
less polarized \citep{Vaillancourt 2002}. On the other hand, we also
find in OMC-1 some of the largest polarization levels ever detected
with Hertz at 350 $\mu$m. For example, the group of dark colored
vectors in the northeastern part of Figure \ref{cap:omc1} have levels
that can reach as high as $\approx10$\%. These levels are found in
a region of relatively low flux where the depolarization effect is
minimal and will be useful for our measurement of the angle of inclination
of the magnetic field that will soon follow.

\subsubsection{The OMC-4 region}

Finally, as far as the polarimetry data are concerned, we show in
Figure \ref{cap:omc4} the Hertz polarization vectors superposed on
the SHARC II continuum map of OMC-4. It is located some $13\arcmin$
south of the KL Nebula in OMC-1 and is seen to be of relatively low
intensity ($\approx10$ Jy in $12\arcsec$). An average of the Stokes
parameters for a few vectors gives a polarization level of $\approx1.5$\%
and a polarization angle of $\approx165^{\circ}$. Although the amount
of polarization measured is similar to what was found in IRAS 05327-457
and OMC-3 MMS 1-6, the orientation of the vectors exceeds what is
found there by some $20^{\circ}$ to $30^{\circ}$. Moreover, their
orientation does not follow the general orientation of the local filament
which makes an angle of approximately $25^{\circ}$.

\subsection{The spectroscopic data - the inclination angle of the magnetic field}

In this section we will make use of the technique put forth by \citet{Houde 2002}
that allows for a measurement of the inclination of the magnetic field
relative to the line of sight. We present in Figure \ref{cap:spectra}
typical spectra (although see section \ref{sub:outflows}) taken in
four of the five regions where we have sought and obtained spectroscopic
data: the northeastern part of OMC-1, and the neighborhoods of OMC-3
MMS 6, OMC-3 MMS 9 and OMC-2 FIR 4. The technique relies on comparisons
of line profiles of coexistent neutral and ion molecular species and
we have chosen the $J\rightarrow4-3$ transition of HCN and HCO$^{+}$,
respectively. The telescope beam width was $\approx20\arcsec$ for
these observations, similar to that obtained while observing with
Hertz.

\subsubsection{Calibration of the ion-to-neutral line width ratio and the calculation
of the inclination angle\label{sub:Calibration}}

As was explained by \citet{Houde 2002}, the evaluation of the inclination
angle of the magnetic field necessitates the combination of spectroscopic
and polarimetry data. This is done by effectively calibrating the
ion-to-neutral line width ratio, calculated a priori from the HCO$^{+}$
and HCN spectra, with the corresponding polarimetry data. In doing
so, one tries to determine the most suitable of a family of curves,
pertaining to the propensity of alignment between the magnetic field
and neutral flows, to apply to the region under study (see the models
of Figures 2, 4 and 5 of \citet{Houde 2002}). For the present case
of Orion A, ideally one would like to determine such a curve for each
region where one desires to locally evaluate the inclination angle
(which will be labeled $\alpha$, along with $\beta$ for the angle
made by the projection of the magnetic field on the plane of the sky).
Unfortunately, the strong amount of depolarization encountered in
most of the regions studied here renders this task impossible. This
can be asserted from Figure \ref{cap:depol} where we plotted the
polarization level against the 350 $\mu$m continuum flux for the
polarimetry data shown for the OMC-3 and OMC-2 fields of Figures \ref{cap:omc3}
and \ref{cap:omc2}, respectively. Since our spectroscopic data had
to be taken in locations of high enough intensity, the corresponding
polarimetry points are all significantly affected by the depolarization
effect.

Faced with this, one could reasonably choose an arbitrary curve corresponding
to the aforementioned set of models (defined by the $\Delta\theta$
parameter of equation (11) of \citet{Houde 2002}, which quantifies
the level of collimation of the neutral flows to the magnetic field).
Although this would bring some uncertainty in the evaluation of $\alpha$,
it should give a reasonably good estimate. This is especially true
if $\alpha$ covers a set of values that are close to the lower and
upper limits of its available range (i.e., $0^{\circ}$ and $90^{\circ}$,
respectively; see the discussion in section 5 of \citet{Houde 2002}).
There was, however, another option available to us. Although the OMC-1
field is also severely depolarized at places, there exists regions
away from the core where the flux is high enough and the depolarization
weak enough to allow us to use the polarization levels measured there
to calibrate the data for all of Orion A. In fact, as stated before,
OMC-1 presents us with some of the highest polarization levels ever
measured with Hertz. We, therefore, chose for this seven such points
for which the corresponding polarization vectors were plotted in a
darker color in Figure \ref{cap:omc1} (around $\Delta\alpha\simeq\mathrm{1.5}\arcmin$
and $\Delta\delta\simeq\mathrm{-7}\arcmin$). The result of the calibration
is shown in Figure \ref{cap:rvsp} where we have set the maximum level
of polarization $P_{max}$ to $10\%$. \citet{Houde 2002} had chosen
$P_{max}=7\%$ in their study of the magnetic field in M17, but we
obviously had to update this parameter in view of our more recent
data. This parameter, just as $\Delta\theta$, has been shown not
to have a significant impact on the results. The model that best fits
our data, using a non-linear least-squares technique, has $\Delta\theta=28.3^{\circ}$
as shown in Figure \ref{cap:rvsp}.

This analysis allows us to give a set of average values for $\alpha$
in the five regions studied here. More precisely, we obtain from simple
averages $\alpha\simeq72.6^{\circ}\pm4.4^{\circ}$, $\simeq73.7^{\circ}\pm5.2^{\circ}$,
$\simeq79.8^{\circ}\pm4.0^{\circ}$, $\simeq65.1^{\circ}\pm9.9^{\circ}$
and $\simeq49.1^{\circ}\pm8.7^{\circ}$ for OMC-3 MMS 6, OMC-3 MMS
9, OMC-2 FIR 4, OMC-1 (northeast), and the Bar, respectively. One
should not give too much significance in the dispersions quoted above
in view of the small numbers of points available in each region. Moreover,
the errors thus calculated result from the uncertainties present in
the fit for the spectra only and do not take into account the uncertainty
in the selection of the model for Orion A shown in Figure \ref{cap:rvsp}
and discussed above. However, because of the large values calculated
for the inclination angle, the latter could amount to at most a few
degrees in the majority of cases. 

\subsubsection{The origin and nature of the neutral flows and the ion line narrowing effect}

Our capacity to evaluate the inclination of the magnetic field to
the line of sight rests almost entirely on the effect originally discussed
by \citet{Houde 2000a} concerning the narrowing of the line profiles
of molecular ions when compared to coexistent neutral species. In
turn, it is believed that this effect depends on the existence of
turbulence in the regions studied; more precisely, the presence of
turbulence in the neutral component of the gas (which we model through
the existence of a large number of neutral flows). It has also long
been assumed that this same turbulence is at the origin of the fact
that the line profiles observed in molecular clouds, like in Orion
A, are usually much broader than their thermal width \citep{Zuckerman 1974,Falgarone 1990}.
The division of a given line width into thermal and non-thermal parts
shows that, for clouds of sufficient mass or luminosity, the later
quickly becomes the dominant component (see for example \citet{Myers 1991}).
Although this turbulent behavior is implied and fully taken advantage
of in the present and previous analyses \citep{Houde 2002, Lai 2003},
the causes and origins of the neutral flows postulated therein have
not been discussed.

Given the different scales encountered in molecular clouds (i.e.,
velocities, spatial dimensions) and the fact that a flow will become
turbulent at large Reynolds numbers, a turbulent regime should easily
be established in a cloud whenever there are bulk motions of matter.
One can, therefore, think of or propose different physical mechanisms
through which this can happen. Perhaps one of the most natural way
of inducing bulk motions is through gravitational interaction. This
could happen, for example, in the case of a cloud or a fragment of
cloud that has reached supercriticality and where the neutral component
of the gas is allowed to collapse through the magnetic field \citep{Mouschovias 1999}.
There is, in fact, observational evidence of supercriticality for many molecular
clouds \citep{Crutcher 1999}. Another mechanism can involve the presence
of protostellar (bipolar) or stellar outflows by which the (mostly
neutral) circumstellar environment is stirred and entrained. High
velocity outflows can also be the source of shocks that can drive
the neutrals through the magnetic field and the ions. Alternatively,
it is possible that the source for the turbulence of the neutral component
of the gas originates outside from its immediate environment. For example,
a neighboring ionized (H {\small II}) region harboring violent magnetohydrodynamic
processes (e.g., Alfvén wave generated winds) could, at the frontier
linking the two regions, mechanically transmit a portion of its turbulent
energy to the neutral gas of an adjacent molecular cloud. It is likely
that other mechanisms exist and can help in accounting for the presence
of the turbulence in the neutral component of the gas necessary to
explain the large line profiles of the neutral molecular species.
The few examples presented here are all likely to play a role in this.
As far as the Orion A complex is concerned, the multiple outflow nature
of this region is well known \citep{Williams 2003} and the combination
of turbulent flows, shocks and stellar outflows at various directions
and speeds projected on the line of sight will sum to a line shape
for neutrals of considerable width. 

The turbulence (and neutral flows) resulting from the aforementioned
physical mechanisms will compete with the local magnetic field in
dictating the dynamics of the ions. A comparison of the strength of
magnetic force acting on the ions with that of the friction force
they are also subjected to (resulting from their collisions with the
particles composing the neutral gas) indicates that the ions are effectively
trapped by the magnetic field \citep{Mouschovias 1999, Houde 2000a}.
This restricts their motion in directions perpendicular to the field
and removes many velocity components from their common observed line
shape that are otherwise present in the spectral profiles of the coexistent
neutral species (that fully take part in the turbulent regime). Moreover,
under a condition of equilibrium where an equipartition of energy
has been attained between the colliding partners (i.e., ions and neutrals),
it is found that the mean gyration velocitiy of an ion around a guiding
center will be less than that of a typical neutral flow (it can be
shown to scale as the square root of the ratio of the (smaller) neutral
mass to the ion mass \citep{Houde 2000a, Houde 2000b}). This will
result, in general, in narrower molecular ion line profiles and explain
the significant amount of ambipolar diffusion thus observed \citep{Houde 2002}. 
\subsubsection{A word of caution about outflows\label{sub:outflows}}

In evaluating the line widths of the HCN and HCO$^{+}$ spectra and,
subsequently, their ratio care must be taken that the spectra studied
are obtained in region suitable to this type of analysis. For example,
the line width calculations and the models proposed by \citet{Houde 2002}
(see their equations (9) and (10) and Figure 2) make certain assumptions
concerning the distribution of the neutral flows in velocity space.
One consequence is that, ideally, the spectra should be even around
the mean velocity. Although this will rarely be the case in turbulent
molecular clouds, one should be careful that the departures from this
idealization are not extreme. Such an example is shown in Figure \ref{cap:omc3_outflow}
where we show a pair of HCN and HCO$^{+}$ spectra taken in the OMC-3
MMS 6 region. We can see the clear signature of a one-sided outflow
in the neutral line profile that is, however, missing from the ion
spectrum. Although the presence of unresolved (bipolar) outflows within
the telescope beam is not un-welcome when trying to calculate the
orientation of the magnetic field, it does not make sense, for example,
to discuss differences between line widths of coexistent neutral and
ion species within a resolved outflow. The most that one can do in
such cases is to look for similarities or differences in the line
profiles and draw appropriate conclusions from them \citep{Houde 2001}.
For the case at hand, the inclusion of the outflow in the line width
calculations brings an artificial reduction in the value of the line
width ratio. More precisely, we have found two more positions of similar
occurrences in OMC-3 MMS 6 (none in the other sources) where the presence
of the one-sided outflow lowers the local HCO$^{+}$ to HCN line width
ratio and brings the corresponding mean inclination angle to $82.2^{\circ}$.
In analyzing the data presented in the previous section, we have carefully
removed the effect of the outflow, after fitting the spectra with
Gaussian profiles, to ensure that we are within confines relevant
to the model of \citet{Houde 2002}. After doing so, for the example
shown in Figure \ref{cap:omc3_outflow}, the line width ratio jumped
from 0.34 to 0.87. This new ratio is very similar to that obtained
in OMC-3 MMS 9, a region which exhibits line profiles that are alike
to those found for OMC-3 MMS 6 when no outflows are present (see Figure
\ref{cap:spectra}). Moreover, it is seen that this apparently significant
change has a relatively mild effect on the value of the inclination
angle; as stated before, the latter has a revised value of $72.6^{\circ}$.
This is not surprising considering the fact that the field appears
to lie close to the plane of the sky where $\alpha$ is quite insensitive
to such changes. No matter which value for the ratio is used, the
conclusion is still that the field is severely inclined in relation
to the line of sight.

\subsubsection{An early assessment of our $\alpha$-measuring technique}

Without the existence of an independent method for measuring the inclination
angle of the magnetic field, it is difficult to determine the level
of success attained using our technique. It is, however, possible
to get an early assessment on its viability by combining the data
presented in this paper with the earlier set for M17 published by
\citet{Houde 2002}. Since M17 exhibits relatively low polarization
levels ($\la4\%$) while Orion A (more precisely OMC-1) covers a range
that goes as high as $\approx10\%$, a combination of both sets of
data will allow us to see how well our technique performs over a large
range of polarization levels. Despite the strong depolarization observed
in regions of higher flux, measurements obtained on the edges of the
clouds where we find the most significant levels of polarization,
should trace differences in the inclination angle from one cloud to
the other.

We produced in Figure \ref{cap:orion_m17} a graph of the combined
sets of data with the normalized polarization level plotted against
the inclination angle. Also shown is a theoretical curve relating
the two parameters (as would be the case in the absence of depolarization).
As can be seen, there is a good correspondence between low (high)
polarization levels and small (large) inclination angles. The polarization
data for M17 are confined to a range where $\alpha\la60^{\circ}$
whereas the Orion A data set mostly covers $\alpha\ga60^{\circ}$,
with most points being pushed down {}``under'' the curve via the
depolarization effect. Moreover, only three data points are located
to the left of the theoretical curve. This is a good indication that
the vast majority of pairs of polarization/line-width-ratio measurements
fall within a domain that is predicted by our technique. Indications
to the contrary could imply some shortcomings in our model and cast
doubts on its adequacy. The results obtained so far are consistent
with expectations.

\section{Discussion}

The results presented in Table \ref{ta:results} give us a glimpse
into the orientation of the magnetic field in Orion A and can form
the basis for an interpretation of its interaction with its environment.
A few characteristics are easily noticeable upon studying Table \ref{ta:results}.
First, the inclination of the magnetic field varies little as one
proceeds north to south along the ISF, this is especially true north
of OMC-1. Second, the orientation of the magnetic field in the plane
of the sky also shows little variations north of OMC-1 while it significantly
changes direction at and within this high mass cloud. On the other
hand, the orientation of the filament varies by some $60^{\circ}$
in the OMC-3 region alone, and by $\approx80^{\circ}$ overall.

While it is true that in some parts (i.e., IRAS 05327-457, OMC-3 MMS
1-6 and north of OMC-2 FIR 3) the projection of the magnetic field
is almost perpendicular to the local filament, and though this could
be significant in itself \citep{Matthews 2001}, we cannot say that
there is a systematic trend for this. Similarly, there does not seem
to exist a tendency for alignment as this is observed only in OMC-2
FIR 6 (see the last column of Table \ref{ta:results}). Future polarimetry
measurements which could fill the gaps in our coverage might, however,
alter this picture. But for the present, it is perhaps more important
to note that the absolute orientation of the sky-projected magnetic
field changes by a relatively small amount over a large extent. In
fact, from our data we see only one region north of OMC-1 (OMC-2 FIR
3) where there is an important deviation in the value of $\beta$
as compared to the other regions observed in OMC-3 and OMC-2. Neglecting
OMC-2 FIR 3, we find that $\beta$ diminishes smoothly as one goes
southward from IRAS 05327-457 to OMC-2 FIR 6, covering values ranging
from $\approx50^{\circ}$ to $\approx25^{\circ}$. It is only when
we reach the vicinity of the high mass star-forming region of OMC-1
that the orientation of the field varies considerably.

\begin{figure}[htbp]
\epsscale{0.8}
\plotone{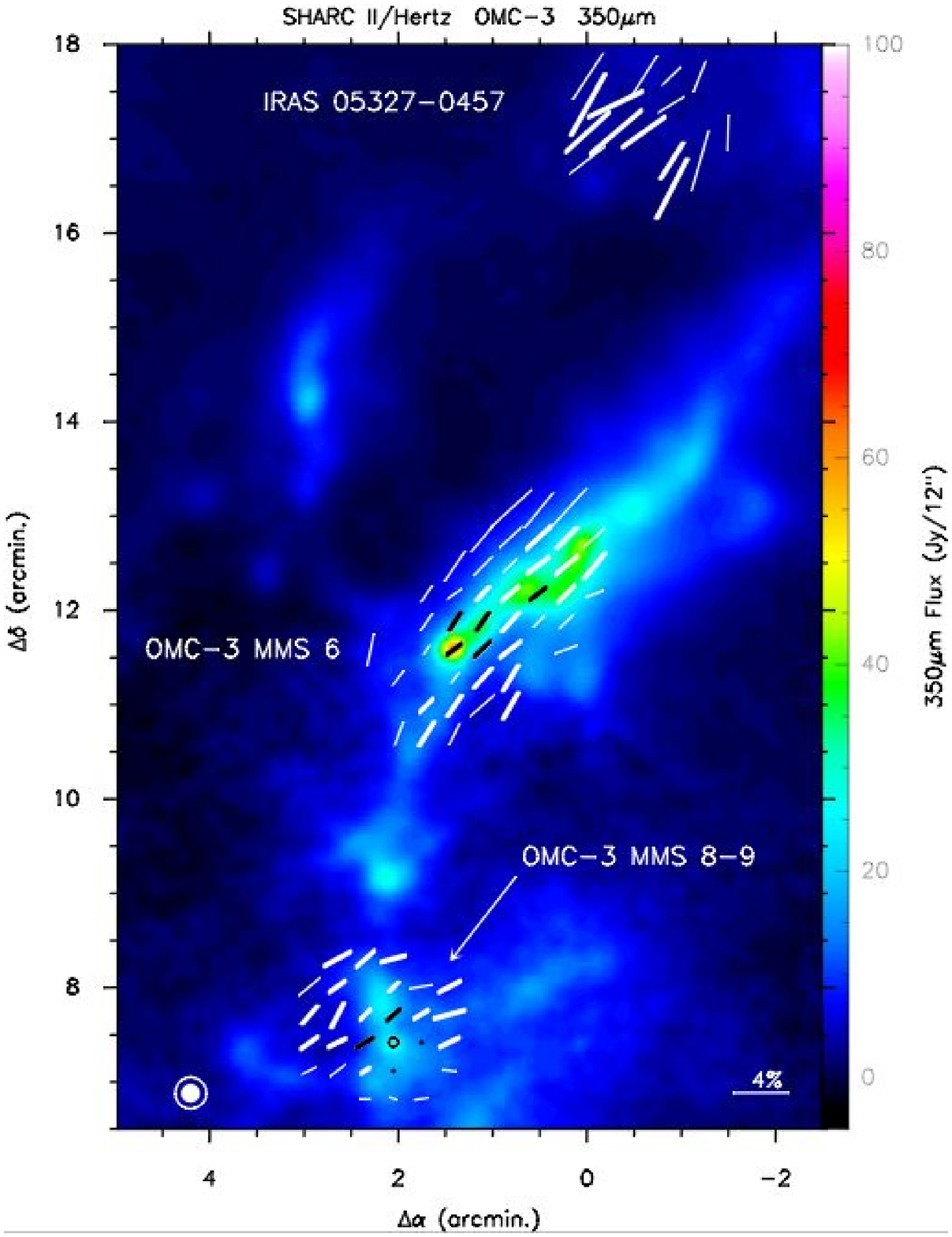}

\caption{\label{cap:omc3}350 $\mu$m continuum map and polarimetry (E-vectors)
of the OMC-3 region obtained with SHARC II and Hertz, respectively.
The thick (thin) vectors have a polarization level and uncertainty
such that $P\geq3\sigma_{P}$ ($P\geq2\sigma_{P}$). The circle indicates
a case where $P+2\sigma_{P}<1\%$ and $P<2\sigma_{P}$. The darker
polarization vectors, circle, and dots denote positions where spectroscopic
data were obtained. The beam widths are shown in the lower left corner,
with the solid and open circles for SHARC II at $\simeq12\arcsec$
and Hertz at $\simeq20\arcsec$, respectively. The reference position
is at R.A.$=5^{h}32^{m}50^{s}$, decl.$=-5\arcdeg15\arcmin00\arcsec$
(B1950).}
\end{figure}

\begin{figure}[htbp]
\epsscale{0.8}
\plotone{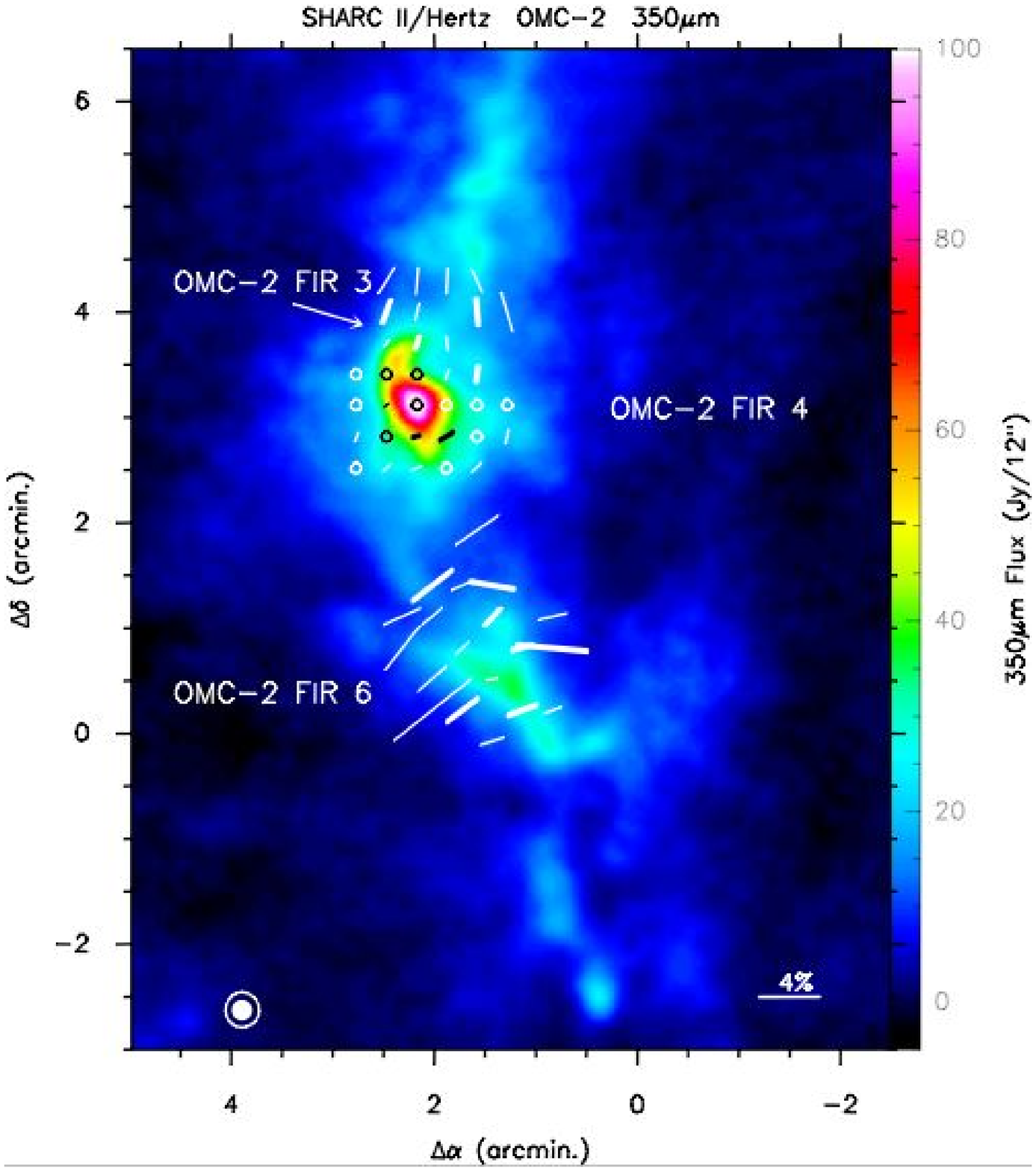}

\caption{\label{cap:omc2}350 $\mu$m continuum map and polarimetry (E-vectors)
of the OMC-2 region obtained with SHARC II and Hertz, respectively.
The thick (thin) vectors have a polarization level and uncertainty
such that $P\geq3\sigma_{P}$ ($P\geq2\sigma_{P}$). Circles indicate
cases where $P+2\sigma_{P}<1\%$ and $P<2\sigma_{P}$. The darker
polarization vectors and circles denote positions where spectroscopic
data were also obtained. The beam widths are shown in the lower left
corner, with the solid and open circles for SHARC II at $\simeq12\arcsec$
and Hertz at $\simeq20\arcsec$, respectively. The reference position
is at R.A.$=5^{h}32^{m}50^{s}$, decl.$=-5\arcdeg15\arcmin00\arcsec$
(B1950). }
\end{figure}

\begin{figure}[htbp]
\epsscale{0.8}
\plotone{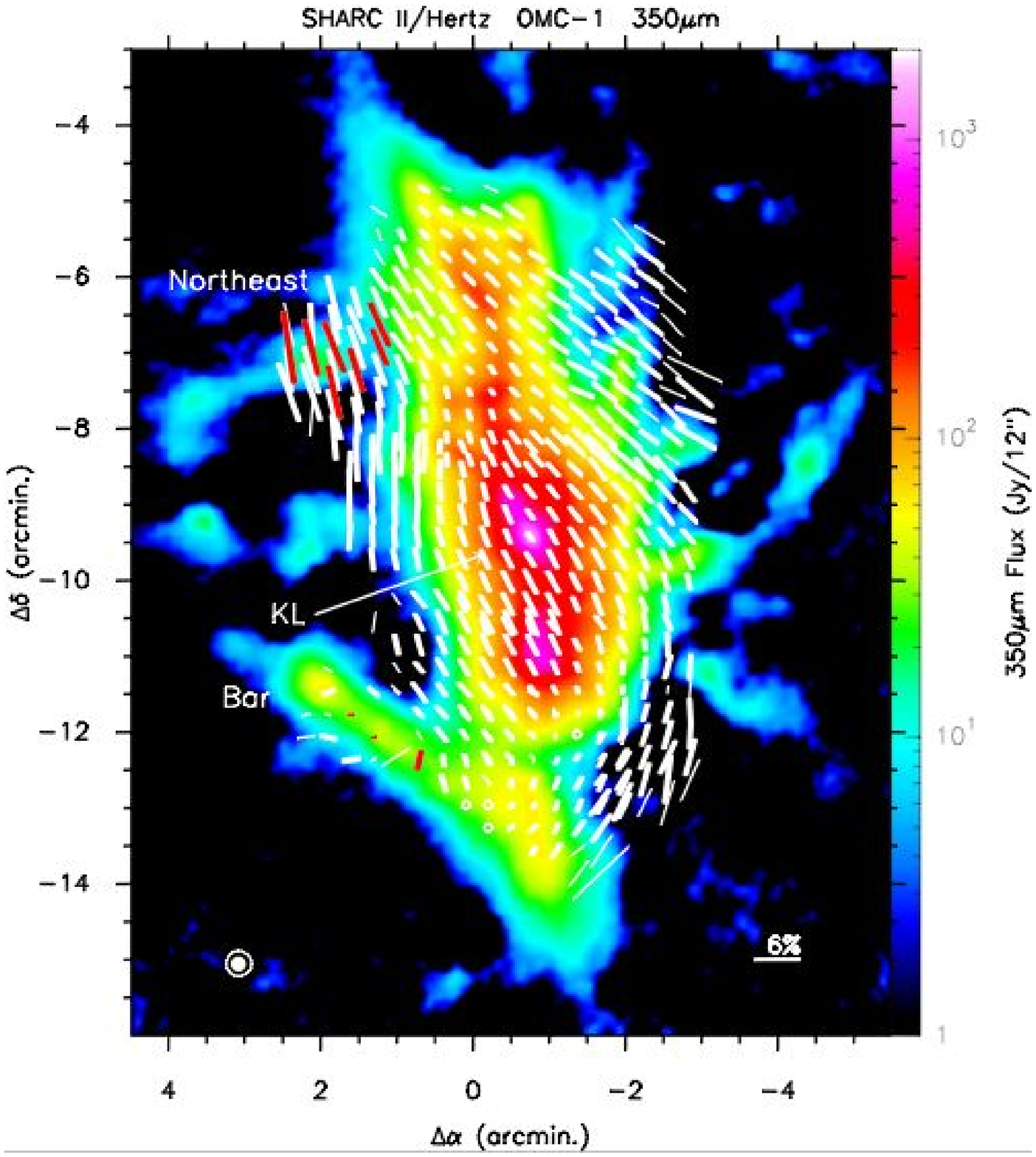}

\caption{\label{cap:omc1}350 $\mu$m continuum map and polarimetry (E-vectors)
of the OMC-1 region obtained with SHARC II and Hertz, respectively.
The thick (thin) vectors have a polarization level and uncertainty
such that $P\geq3\sigma_{P}$ ($P\geq2\sigma_{P}$). Circles indicate
cases where $P+2\sigma_{P}<1\%$ and $P<2\sigma_{P}$. The darker
polarization vectors denote positions where spectroscopic data were
also obtained. The beam widths are shown in the lower left corner,
with the solid and open circles for SHARC II at $\simeq12\arcsec$
and Hertz at $\simeq20\arcsec$, respectively. The reference position
is at R.A.$=5^{h}32^{m}50^{s}$, decl.$=-5\arcdeg15\arcmin00\arcsec$
(B1950). }
\end{figure}

\begin{figure}[htbp]
\epsscale{0.8}
\plotone{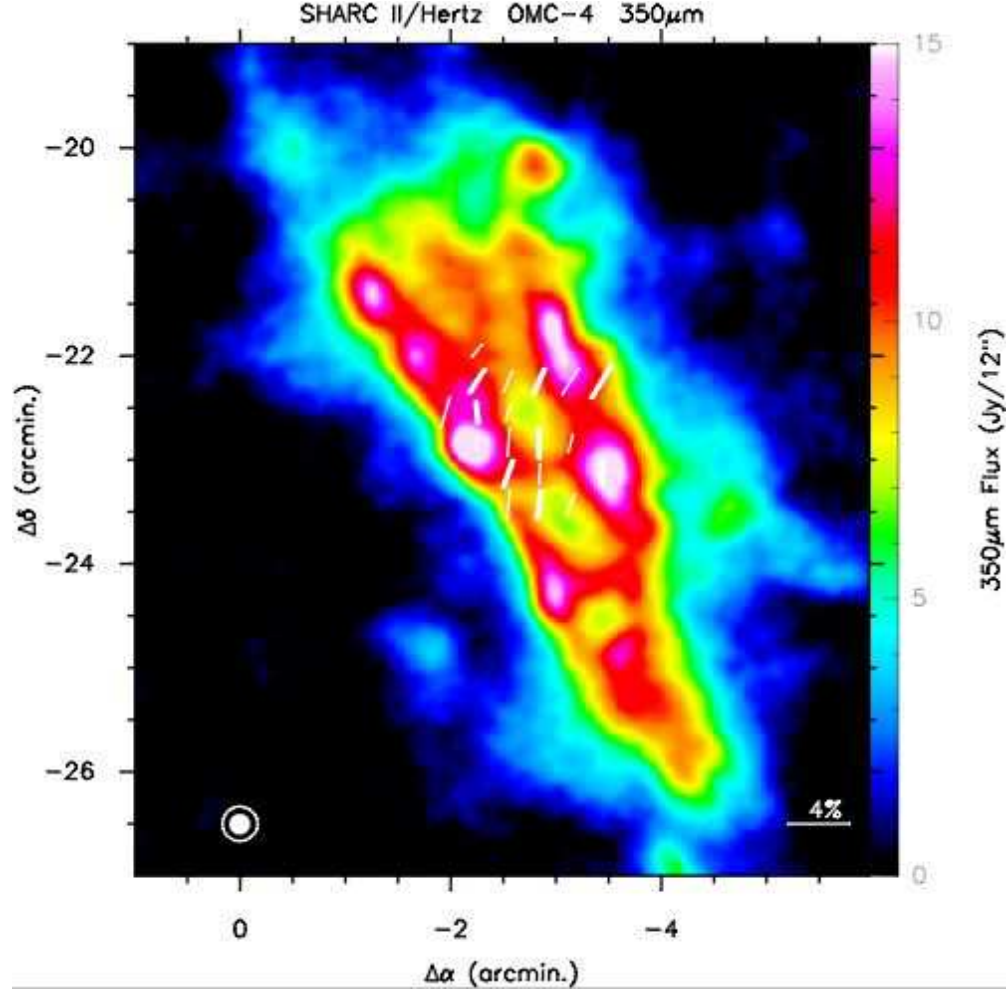}

\caption{\label{cap:omc4}350 $\mu$m continuum map and polarimetry (E-vectors)
of the OMC-4 region obtained with SHARC II and Hertz, respectively.
The thick (thin) vectors have a polarization level and uncertainty
such that $P\geq3\sigma_{P}$ ($P\geq2\sigma_{P}$). The beam widths
are shown in the lower left corner, with the solid and open circles
for SHARC II at $\simeq12\arcsec$ and Hertz at $\simeq20\arcsec$,
respectively. The reference position is at R.A.$=5^{h}32^{m}50^{s}$,
decl.$=-5\arcdeg15\arcmin00\arcsec$ (B1950). }
\end{figure}

\begin{figure}[htbp]
\epsscale{0.8}\plottwo{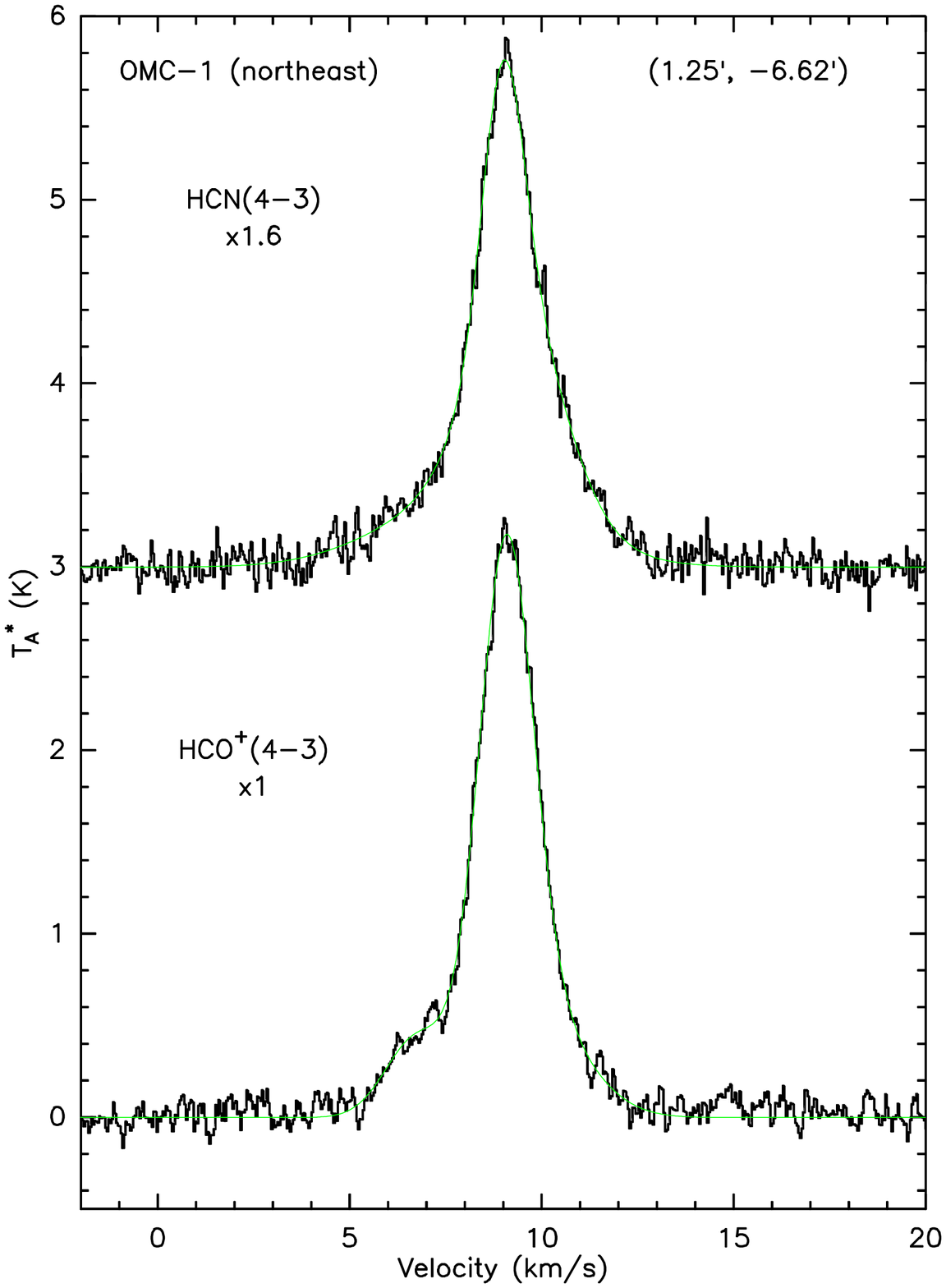}{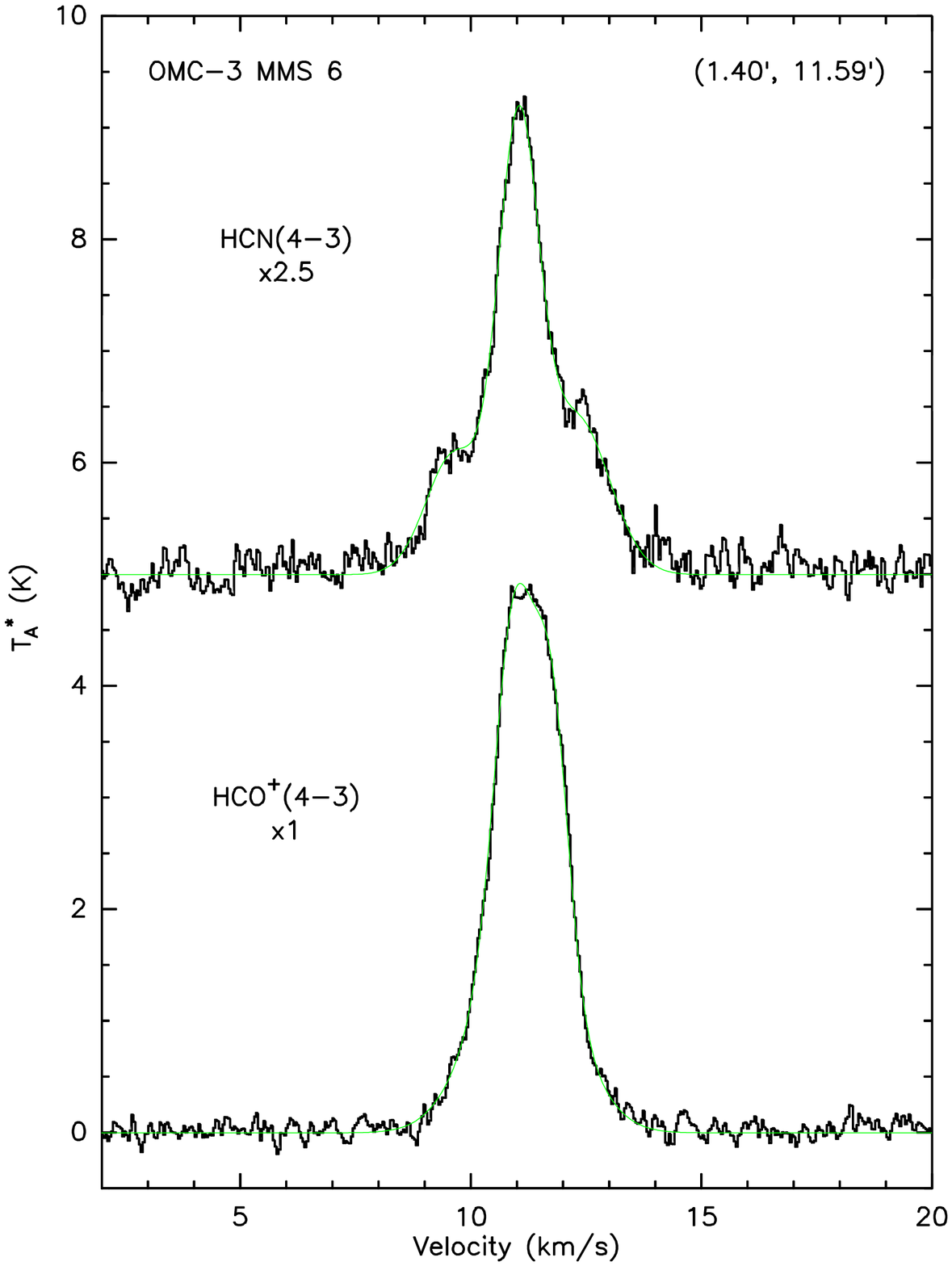} \\

\epsscale{1.8}\plottwo{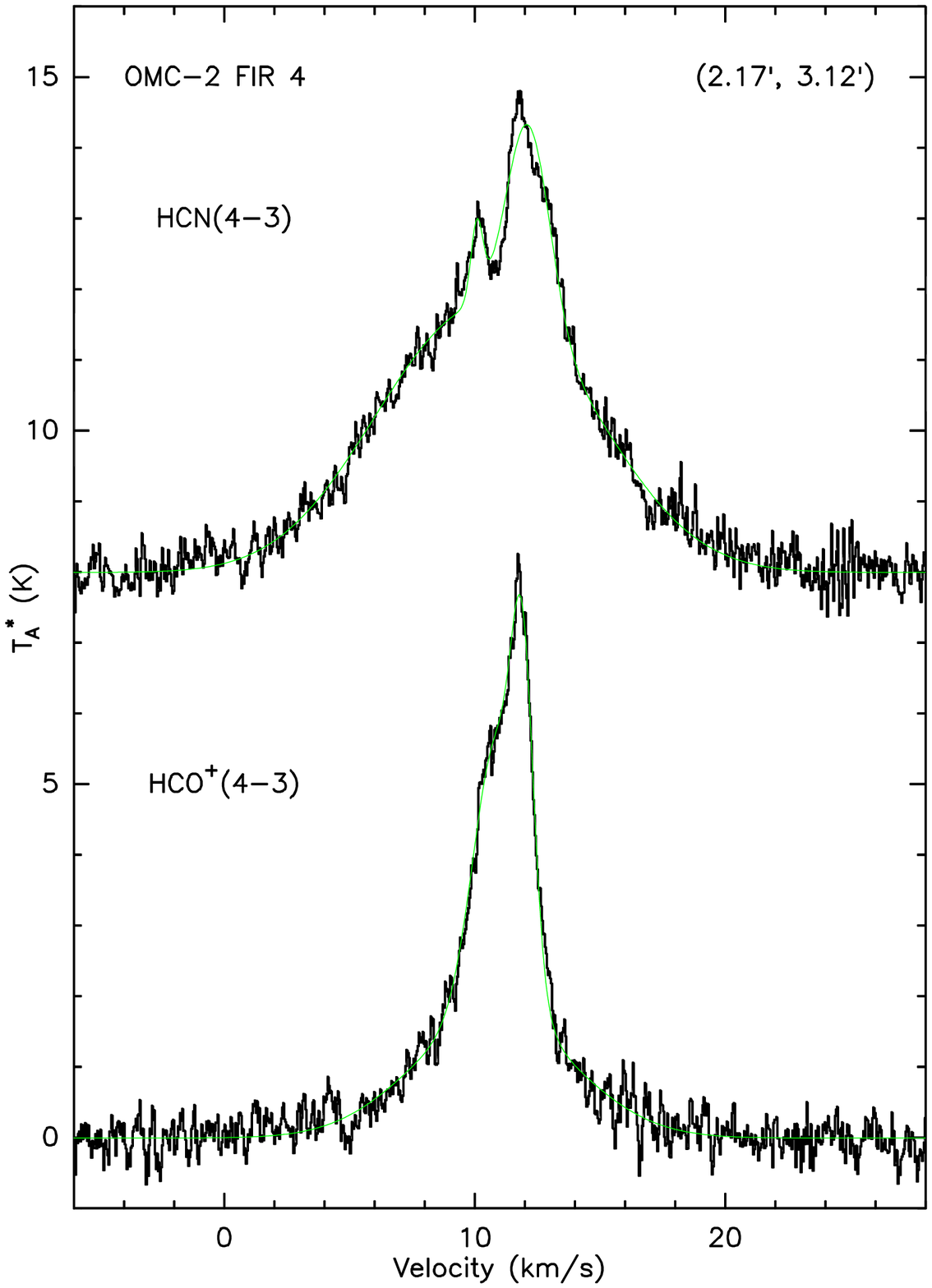}{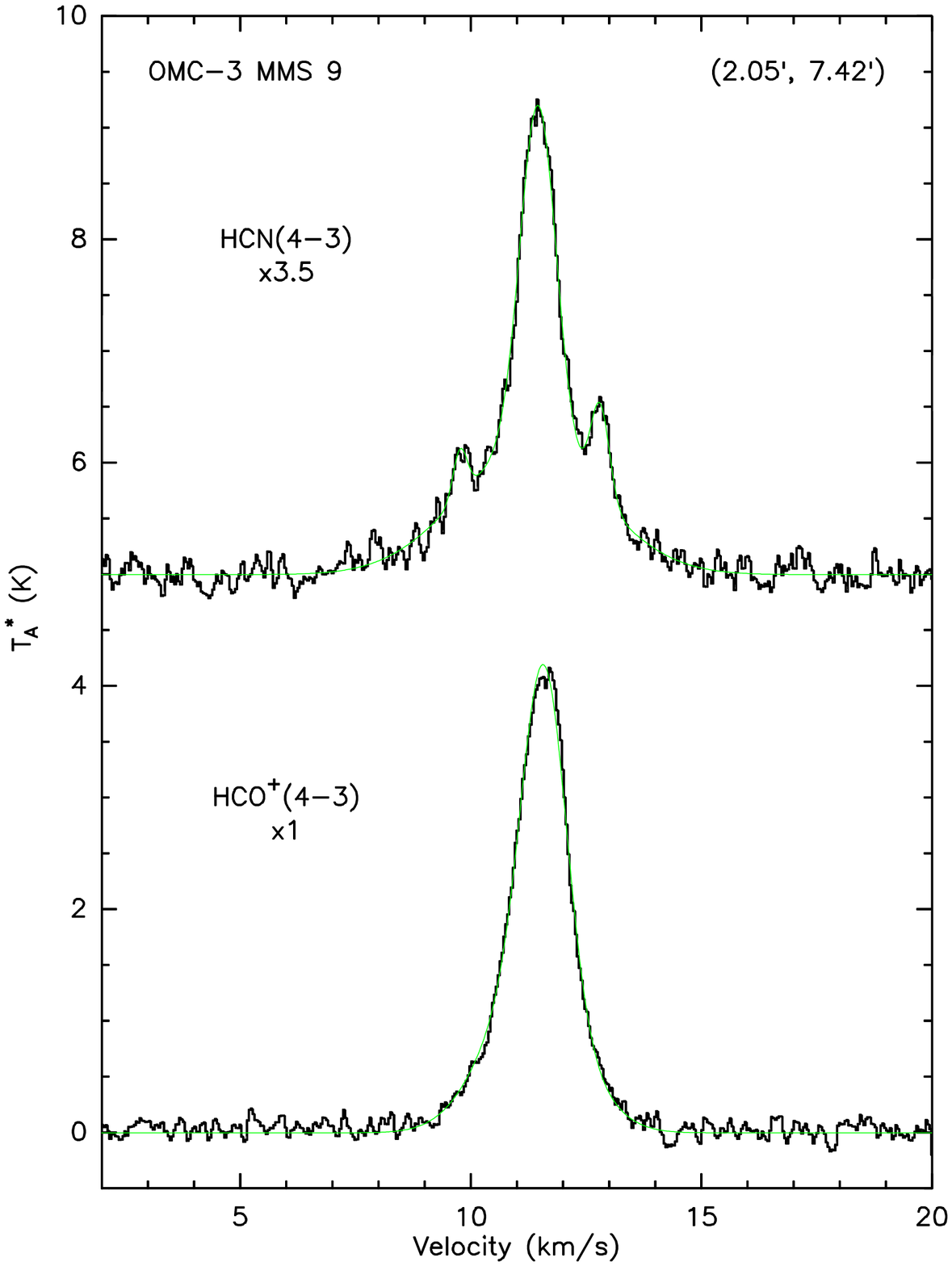}

\caption{\label{cap:spectra}HCN and HCO$^{+}$ spectra in the $J=4\rightarrow3$
transition for, from top-left and clockwise, OMC-1 (northeast), OMC-3
MMS 6, OMC-3 MMS 9 and OMC-2 FIR 4, respectively. The offset position
from the reference at R.A.$=5^{h}32^{m}50^{s}$, decl.$=-5\arcdeg15\arcmin00\arcsec$
(B1950) (see Figure \ref{cap:oriona}) is given in the upper right
corner of each pair of spectra, in arcminutes. }
\end{figure}

\begin{figure}[htbp]
\epsscale{0.6}\rotatebox{270}{
\plotone{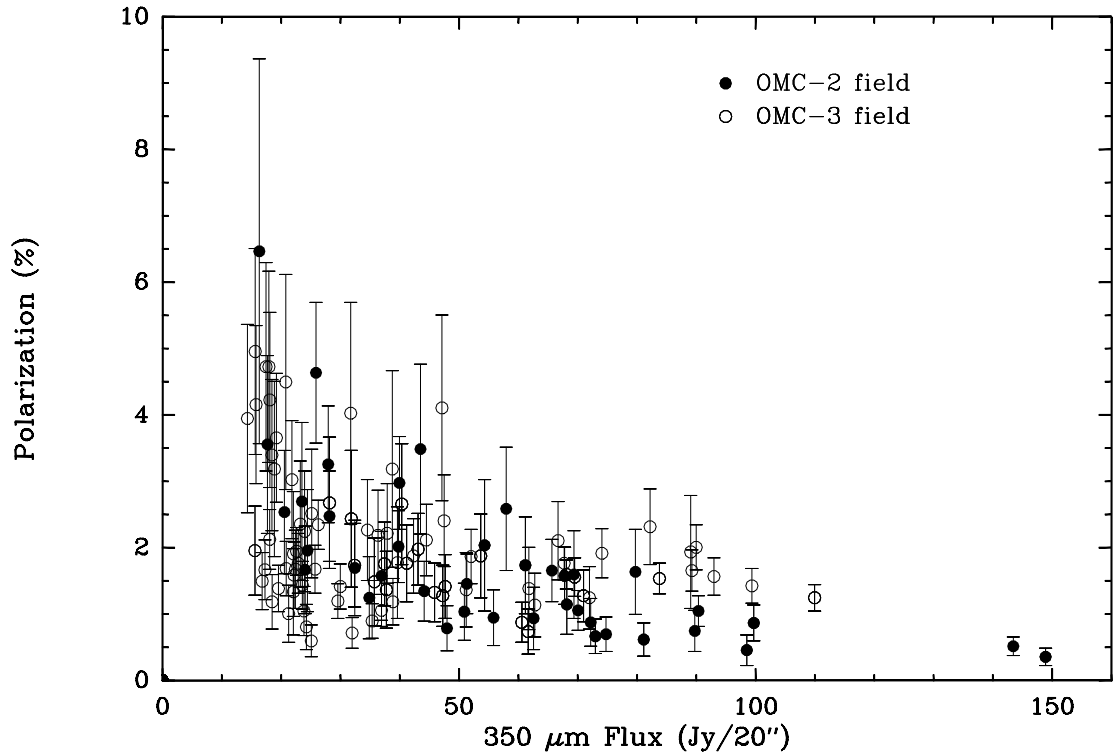}}

\vspace{1cm}

\caption{\label{cap:depol}The polarization level vs the 350 $\mu$m continuum
flux for the polarimetry data as shown for the OMC-3 and OMC-2 fields
of Figures \ref{cap:omc3} and \ref{cap:omc2}, respectively, where
$P\geq2\sigma_{P}$. The depolarization effect is clearly seen.}
\end{figure}

\begin{figure}[htbp]
\epsscale{0.6}\rotatebox{270}{
\plotone{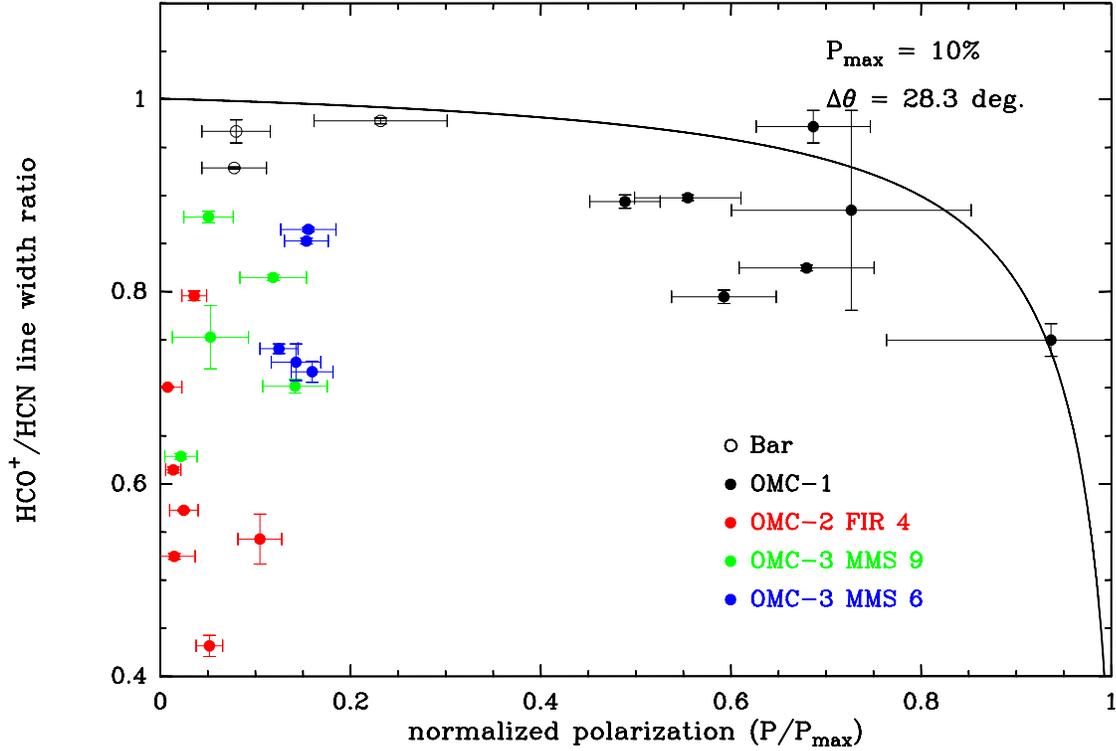}}

\vspace{1cm}

\caption{\label{cap:rvsp}The HCO$^{+}$/HCN line width ratio vs the normalized
polarization level ($P/P_{max}$) for Orion A. $P_{max}$ is set at
10\% and the data are shown against a model of neutral flow collimation
of $\Delta\theta=28.3^{\circ}$(solid curve). The high level of depolarization
in OMC-3 and OMC-2 is clearly seen. The polarization vectors used
to determine the model all have $P>3\sigma_{P}$. }
\end{figure}

\begin{figure}[htbp]
\epsscale{0.7}
\plotone{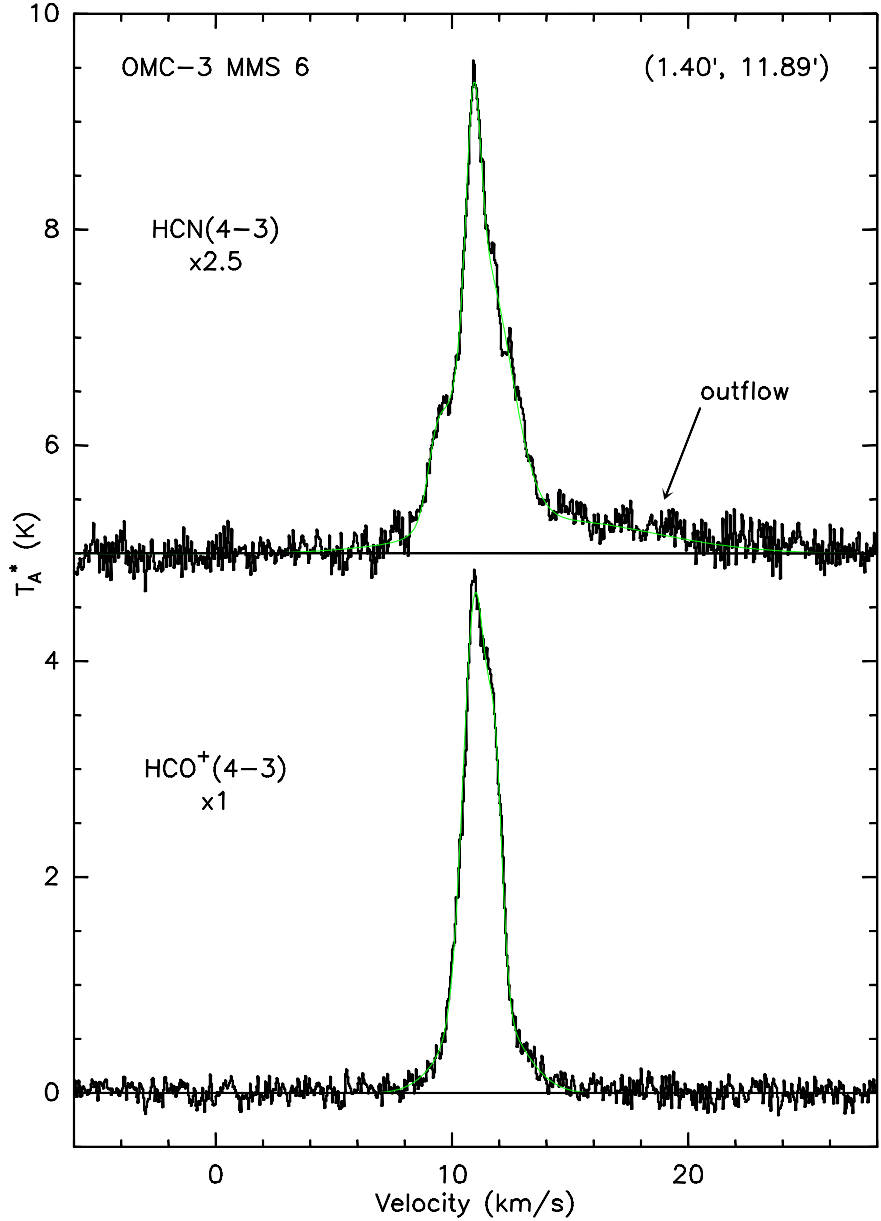}

\caption{\label{cap:omc3_outflow}Spectra taken in OMC-3 MMS 6 that show the
effect of a one-sided outflow on the evaluation of the HCO$^{+}$/HCN
line width ratio. The outflow is detected in HCN but not in HCO$^{+}$.
The line width ratio with and without accounting for the outflow is
0.34 and 0.87, respectively (see the text). The offset position from
the reference at R.A.$=5^{h}32^{m}50^{s}$, decl.$=-5\arcdeg15\arcmin00\arcsec$
(B1950) (see Figure \ref{cap:oriona}) is given in the upper right
corner, in arcminutes. }
\end{figure}

\begin{figure}[htbp]
\epsscale{0.7}
\plotone{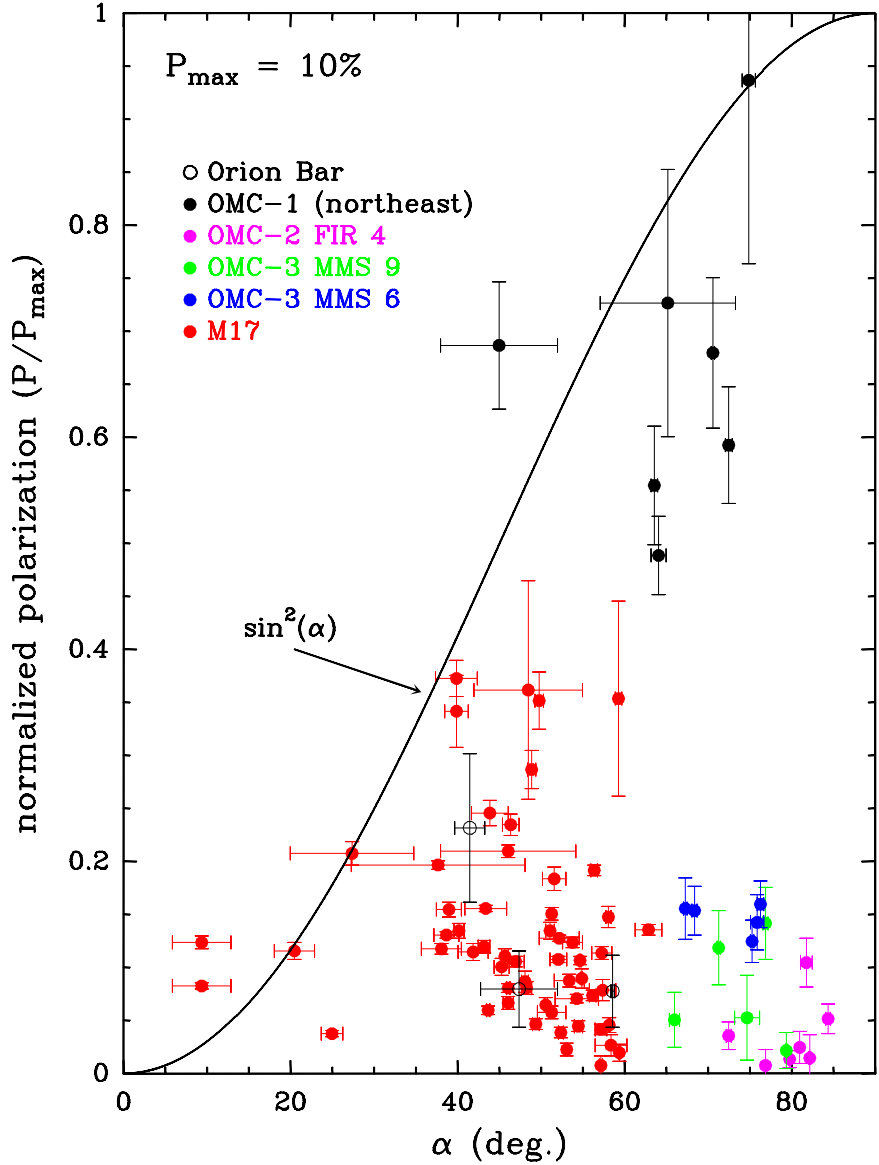}

\caption{\label{cap:orion_m17}Normalized polarization level ($P/P_{max}$)
vs the inclination angle ($\alpha$) for every source studied so far.
The theoretical relation between the two parameters is also shown
(solid curve). There is a good correspondence of low (high) polarization
levels to small (large) inclination angles as would be expected.}
\end{figure}

\begin{figure}[htbp]
\epsscale{0.55}
\plotone{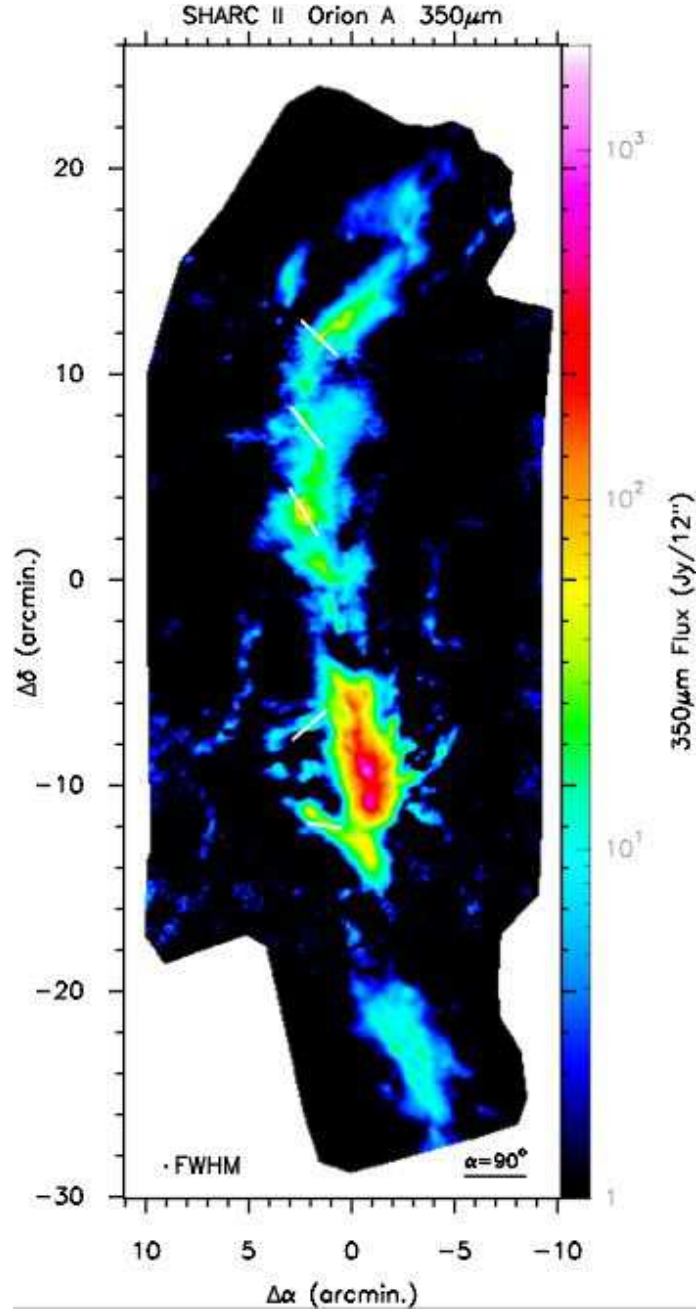}

\caption{\label{cap:oriona}350 $\mu$m continuum map of the Orion A region
obtained with SHARC II. The orientation of the magnetic field is indicated
at five positions along the ISF on the map. The projection of the
magnetic field in the plane of the sky is also shown by the orientation
of the accompanying vectors and the inclination angle is given by
the length of the vectors (using the scale shown in the bottom right
corner). The beam width is shown in the lower left corner ($\simeq12\arcsec$)
and the reference position is at R.A.$=5^{h}32^{m}50^{s}$, decl.$=-5\arcdeg15\arcmin00\arcsec$
(B1950). }
\end{figure}

We could interpret these results as being consistent with a picture
where the magnetic field is relatively unaffected by the presence
of the concentrations of lower mass that characterize the OMC-3 and
OMC-2 fields (Figures \ref{cap:omc3} and \ref{cap:omc2}), adopting
there an orientation similar to that which it may have on the larger
scale. This aspect of the orientation of the magnetic field in connection
to the larger scale will be treated in an upcoming paper by Poidevin
and Bastien (in preparation). Moreover, the small changes in the inclination
of the magnetic field to the line of sight in these parts of Orion
A only reinforces the idea of a relatively unaffected magnetic field.
The only significant variations in the value of $\alpha$ happen in
OMC-1 (northeast) and in the Bar where it decreases to $\simeq65^{\circ}$
and $\simeq49^{\circ}$, respectively, from $\simeq80^{\circ}$ in
OMC-2 FIR 4. This is also accompanied by significant changes in $\beta$.
This, in itself is not surprising for one would expect the magnetic
field to be strongly perturbed by the presence of the large concentrations
of mass found within OMC-1  and the greater amount of recent star
formation in this region. The radiation and strong stellar winds emanating
from the stars of the Trapezium must also affect the local magnetic
field through their impact on the ionization fraction. The orientation
of the magnetic field is indicated for five different positions along
the ISF on the SHARC II map of Figure \ref{cap:oriona}.

Finally, OMC-1 is one of the few clouds where there exists a Zeeman
detection in CN. \citet{Crutcher 1999} reported a line-of-sight magnetic
field strength of 360 $\mu$G at a position $\Delta\alpha\approx10\arcsec$,
$\Delta\delta\approx20\arcsec$ away from IRc 2. If we assume that
our value of $\alpha\simeq65^{\circ}$ obtained in OMC-1 (northeast)
applies equally well at this location, we calculate a magnitude of
$\approx850\,\,\mu$G for the magnetic field. This is not too far
from that which was used by \citet{Crutcher 1999}, based on statistical
arguments, and therefore corroborates the conclusions reached there.
More precisely, OMC-1 is magnetically supercritical with a mass-to-flux
ratio $M/\Phi_{B}\approx2.6$, and magnetic-to-gravitational and kinetic-to-gravitational
energy ratios of $\approx0.26$ and $\approx0.30$, respectively.
The combination of the magnetic and internal motion energies can therefore
provide a significant amount of support against gravitation.

\acknowledgements{}

We thank Min Yang and Attila Kovacs for their assistance in collecting
and analyzing the SHARC II images. The Caltech Submillimeter Observatory
is funded by the NSF through contract AST 9980846 and the observations
made with Hertz were supported by NSF Grants AST 9987441 and AST 0204886.

\clearpage


\clearpage

 \begin{deluxetable}{lrrrrrr}
 \tabletypesize{\footnotesize}
 \tablecaption{Results - Orientation of the magnetic field in Orion A\label{ta:results}}
 \tablecolumns{7}
 \tablewidth{0pt}
 \tablehead{
 \colhead{Object} &
 \colhead{$\alpha$ \tablenotemark{a}} &
 \colhead{$\sigma_{\alpha}$ \tablenotemark{b}} &
 \colhead{$\beta$ \tablenotemark{c}} &
 \colhead{$\sigma_{\beta}$} & 
 \colhead{$\gamma$ \tablenotemark{d}} &
 \colhead{$\mid\beta-\gamma\mid$}}

 \startdata
 IRAS 05327-457 & \nodata & \nodata & $52.6^{\circ}$ & $3.0^{\circ}$ & $\sim132^{\circ}$ & $\sim79^{\circ}$ \\ 
 OMC-3 MMS 6 & $72.6^{\circ}$ & $4.4^{\circ}$ & $46.3^{\circ}$ & $1.0^{\circ}$ & $\sim132^{\circ}$ & $\sim86^{\circ}$ \\ 
 OMC-3 MMS 9 & $73.7^{\circ}$ & $5.2^{\circ}$ & $38.5^{\circ}$ & $2.8^{\circ}$ & $\sim8^{\circ}$ & $\sim31^{\circ}$ \\ 
 OMC-2 FIR 3 & \nodata & \nodata & $84.6^{\circ}$ & $3.8^{\circ}$ & $\sim160^{\circ}$ & $\sim75^{\circ}$ \\ 
 OMC-2 FIR 4 & $79.8^{\circ}$ & $4.0^{\circ}$ & $31.3^{\circ}$ & $4.6^{\circ}$ & $\sim0^{\circ}$ & $\sim31^{\circ}$ \\ 
 OMC-2 FIR 6 & \nodata & \nodata & $25.0^{\circ}$ & $6.2^{\circ}$ & $\sim35^{\circ}$ & $\sim10^{\circ}$ \\ 
 OMC-1 (northeast) & $65.1^{\circ}$ & $9.9^{\circ}$ & $130.2^{\circ}$ & $0.3^{\circ}$ & $\sim10^{\circ}$ & $\sim120^{\circ}$ \\ 
 Bar & $49.1^{\circ}$ & $8.7^{\circ}$ & $81.8^{\circ}$ & $3.6^{\circ}$ & $\sim52^{\circ}$ & $\sim30^{\circ}$ \\ 
 OMC-4 & \nodata & \nodata & $73.7^{\circ}$ & $2.8^{\circ}$ & $\sim25^{\circ}$ & $\sim49^{\circ}$ \\ 
 \enddata

\tablenotetext{a} {Mean inclination angle of the magnetic field from the line of sight.}
\tablenotetext{b} {Does not included errors in collimation model (see text).}
 \tablenotetext{c} {Stokes average of the position angle of the projection of the magnetic field in the plane of the sky, east from north.}
 \tablenotetext{d} {Approximate position angle of the filament, east of north.}

 \end{deluxetable}

%%% Local Variables: 
%%% mode: latex
%%% TeX-master: "~/Documents/sharcii/omc3.graphic"
%%% End: 

 \begin{deluxetable}{rrrrrrr}
 \tabletypesize{\footnotesize}
 \tablecaption{OMC-1, 350 \micron\ Results\label{ta:omc1}}
 \tablecolumns{7}
 \tablewidth{0pt}
 \tablehead{
 \colhead{$\Delta\alpha$ \tablenotemark{a}} &
 \colhead{$\Delta\delta$ \tablenotemark{a}} &
 \colhead{$P$} & \colhead{$\sigma_{P}$} &
 \colhead{$PA$ \tablenotemark{b}} & \colhead{$\sigma_{PA}$} &
 \colhead{Flux \tablenotemark{c}}}

 \startdata
 -1.69 & -13.85 & 10.17 &  5.02 & 133.6 &  14.1 &   54.2 \\ 
 -1.39 & -13.85 &  4.08 &  1.40 & 143.9 &   9.7 &   79.7 \\ 
 -1.69 & -13.56 &  5.09 &  2.06 & 147.1 &  11.4 &   54.2 \\ 
 -1.39 & -13.56 &  2.33 &  1.08 & 125.4 &  13.3 &   80.4 \\ 
 -1.09 & -13.56 &  2.04 &  0.47 & 148.3 &   6.7 &  115.7 \\ 
 -0.80 & -13.56 &  1.11 &  0.29 & 154.8 &   7.5 &  139.8 \\ 
 -1.69 & -13.26 &  4.22 &  1.11 & 146.2 &   6.4 &   57.9 \\ 
 -1.39 & -13.26 &  2.18 &  0.47 & 147.2 &   6.0 &   79.6 \\ 
 -1.09 & -13.26 &  1.22 &  0.33 & 157.4 &   7.8 &  109.9 \\ 
 -0.80 & -13.26 &  1.09 &  0.17 & 139.7 &   4.1 &  141.6 \\ 
 -0.50 & -13.26 &  0.82 &  0.22 & 120.8 &   7.6 &  134.1 \\ 
 -0.20 & -13.26 &  0.22 &  0.24 &  95.0 &  31.2 &  110.1 \\ 
 -1.98 & -12.96 &  3.47 &  1.04 & 153.3 &   8.5 &   53.1 \\ 
 -1.69 & -12.96 &  2.28 &  0.73 & 154.3 &   9.2 &   61.5 \\ 
 -1.39 & -12.96 &  1.66 &  0.36 & 161.8 &   6.3 &   77.7 \\ 
 -1.09 & -12.96 &  1.14 &  0.20 & 162.1 &   5.1 &  104.7 \\ 
 -0.80 & -12.96 &  0.63 &  0.21 & 158.9 &   9.2 &  145.6 \\ 
 -0.50 & -12.96 &  0.63 &  0.20 & 141.4 &   6.2 &  141.4 \\ 
 -0.20 & -12.96 &  0.14 &  0.19 & 113.9 &  39.3 &  129.7 \\ 
  0.09 & -12.96 &  0.33 &  0.23 &  56.8 &  20.1 &  106.0 \\ 
 -2.56 & -12.92 &  6.17 &  2.15 & 163.2 &  10.1 &   29.8 \\ 
 -2.26 & -12.92 &  4.03 &  1.77 & 155.0 &  12.6 &   37.4 \\ 
 -1.96 & -12.92 &  5.25 &  1.11 & 154.2 &   6.1 &   53.5 \\ 
 -1.67 & -12.92 &  3.94 &  0.74 & 143.9 &   5.4 &   76.7 \\ 
 -1.98 & -12.67 &  3.19 &  1.04 & 155.8 &   9.4 &   52.6 \\ 
 -1.69 & -12.67 &  2.06 &  0.66 & 157.1 &   9.2 &   71.3 \\ 
 -1.39 & -12.67 &  1.78 &  0.32 & 151.9 &   5.2 &   80.7 \\ 
 -1.09 & -12.67 &  1.60 &  0.21 & 161.1 &   3.7 &   97.4 \\ 
 -0.80 & -12.67 &  1.24 &  0.14 & 168.3 &   3.4 &  133.8 \\ 
 -0.20 & -12.67 &  0.96 &  0.33 &   2.3 &   5.9 &  133.5 \\ 
 -0.50 & -12.67 &  0.90 &  0.15 & 179.2 &   4.0 &  142.6 \\ 
  0.09 & -12.67 &  1.61 &  0.24 &  14.7 &   4.2 &  124.4 \\ 
  0.39 & -12.67 &  2.18 &  0.36 &  16.7 &   4.7 &  110.8 \\ 
 -2.85 & -12.62 &  6.09 &  2.55 & 156.3 &  12.0 &   26.9 \\ 
 -2.56 & -12.62 &  6.13 &  1.22 & 173.5 &   5.9 &   36.4 \\ 
 -2.26 & -12.62 &  4.38 &  0.91 & 165.5 &   6.0 &   46.5 \\ 
 -1.96 & -12.62 &  2.98 &  0.59 & 149.0 &   5.8 &   67.9 \\ 
 -1.67 & -12.62 &  1.51 &  0.49 & 145.8 &   9.5 &   92.1 \\ 
 -0.18 & -12.62 &  1.54 &  0.76 &  41.4 &  15.1 &  161.9 \\ 
 -1.98 & -12.37 &  2.71 &  0.68 & 148.0 &   7.0 &   65.2 \\ 
 -1.69 & -12.37 &  1.77 &  0.44 & 152.4 &   7.2 &   81.9 \\ 
 -1.09 & -12.37 &  0.65 &  0.20 & 152.4 &   8.8 &  115.1 \\ 
 -0.80 & -12.37 &  0.40 &  0.17 & 157.3 &  12.7 &  131.6 \\ 
 -0.50 & -12.37 &  0.77 &  0.15 &   7.4 &   5.5 &  138.4 \\ 
 -0.20 & -12.37 &  1.15 &  0.22 &  20.4 &   5.4 &  141.0 \\ 
  0.09 & -12.37 &  1.38 &  0.24 &  14.2 &   4.9 &  125.5 \\ 
  0.39 & -12.37 &  1.41 &  0.33 &   5.0 &   6.7 &  114.8 \\ 
  0.71 & -12.36 &  2.32 &  0.70 & 167.6 &   8.6 &  106.6 \\ 
  1.30 & -12.36 &  0.88 &  0.37 & 115.5 &  12.1 &   87.9 \\ 
  1.60 & -12.36 &  2.19 &  0.68 &  96.3 &   8.8 &   61.5 \\ 
 -2.85 & -12.32 &  4.72 &  1.36 & 165.5 &   8.4 &   36.1 \\ 
 -2.56 & -12.32 &  4.51 &  0.92 & 154.4 &   5.8 &   57.3 \\ 
 -2.26 & -12.32 &  3.12 &  0.61 & 159.1 &   5.6 &   71.5 \\ 
 -1.96 & -12.32 &  2.53 &  0.42 & 150.2 &   4.8 &  102.2 \\ 
 -1.67 & -12.32 &  0.89 &  0.37 & 142.3 &  11.1 &  136.2 \\ 
 -1.37 & -12.32 &  0.80 &  0.24 & 146.3 &   8.4 &  167.3 \\ 
 -0.78 & -12.32 &  0.72 &  0.28 &  59.0 &  11.3 &  213.8 \\ 
 -0.48 & -12.32 &  0.98 &  0.23 &  50.8 &   6.8 &  198.3 \\ 
 -0.18 & -12.32 &  1.52 &  0.34 &  30.7 &   7.3 &  173.7 \\ 
  0.11 & -12.32 &  1.42 &  0.62 &  41.2 &   9.2 &  139.7 \\ 
  1.00 & -12.32 &  4.41 &  2.02 & 122.9 &  13.0 &  145.6 \\ 
 -1.98 & -12.07 &  1.86 &  0.66 & 155.7 &  10.2 &   87.8 \\ 
 -1.69 & -12.07 &  1.44 &  0.39 & 147.9 &   7.9 &  112.2 \\ 
 -1.09 & -12.07 &  0.51 &  0.17 & 138.4 &   9.4 &  153.3 \\ 
 -0.80 & -12.07 &  0.40 &  0.14 &  79.6 &  10.2 &  175.1 \\ 
 -0.50 & -12.07 &  0.60 &  0.13 &  51.6 &   7.3 &  164.4 \\ 
 -0.20 & -12.07 &  0.87 &  0.19 &  30.5 &   6.2 &  146.1 \\ 
  0.09 & -12.07 &  0.71 &  0.25 &  40.5 &   9.9 &  131.3 \\ 
  0.39 & -12.07 &  1.00 &  0.42 &  17.4 &  11.9 &  122.4 \\ 
  2.19 & -12.07 &  2.42 &  0.82 &  97.6 &   9.6 &   49.1 \\ 
  1.30 & -12.07 &  0.80 &  0.36 & 104.9 &  12.9 &  117.3 \\ 
  1.89 & -12.07 &  1.97 &  0.55 &  77.9 &   8.0 &   66.4 \\ 
 -2.85 & -12.03 &  3.51 &  0.97 & 177.3 &   8.0 &   58.3 \\ 
 -2.56 & -12.03 &  3.65 &  0.99 & 166.9 &   6.1 &   69.7 \\ 
 -2.26 & -12.03 &  2.72 &  0.44 & 162.3 &   4.6 &   89.2 \\ 
 -1.96 & -12.03 &  1.65 &  0.39 & 160.0 &   7.2 &  133.2 \\ 
 -1.67 & -12.03 &  0.62 &  0.19 & 167.8 &   9.0 &  200.7 \\ 
 -1.37 & -12.03 &  0.01 &  0.15 & 115.9 & 452.5 &  244.7 \\ 
 -1.07 & -12.03 &  0.65 &  0.25 & 150.6 &  13.8 &  285.8 \\ 
 -0.78 & -12.03 &  0.77 &  0.17 &  38.5 &   5.6 &  310.5 \\ 
 -0.48 & -12.03 &  1.47 &  0.21 &  43.7 &   3.2 &  249.4 \\ 
 -0.18 & -12.03 &  1.33 &  0.19 &  40.3 &   4.1 &  188.5 \\ 
  0.11 & -12.03 &  1.36 &  0.27 &  32.6 &   5.8 &  162.4 \\ 
  0.41 & -12.03 &  2.17 &  0.47 &  36.5 &   6.1 &  120.6 \\ 
  0.71 & -12.03 &  1.31 &  0.60 &  24.8 &  13.1 &  130.8 \\ 
 -1.98 & -11.78 &  4.64 &  1.39 & 176.5 &   8.7 &  107.3 \\ 
 -1.69 & -11.78 &  1.00 &  0.34 & 146.9 &  10.1 &  167.4 \\ 
 -1.39 & -11.78 &  0.63 &  0.25 & 123.1 &  11.0 &  200.9 \\ 
 -1.09 & -11.78 &  0.58 &  0.15 & 102.7 &   7.0 &  234.3 \\ 
 -0.80 & -11.78 &  0.81 &  0.13 &  52.3 &   6.9 &  241.0 \\ 
 -0.50 & -11.78 &  1.48 &  0.17 &  45.0 &   4.0 &  217.9 \\ 
 -0.20 & -11.78 &  1.42 &  0.23 &  29.6 &   4.6 &  171.9 \\ 
  0.09 & -11.78 &  1.41 &  0.39 &  44.1 &   7.9 &  147.6 \\ 
  0.71 & -11.77 &  1.09 &  0.51 &  54.3 &  13.4 &   83.9 \\ 
  1.60 & -11.77 &  0.78 &  0.34 &  85.2 &  12.4 &  108.5 \\ 
  1.89 & -11.77 &  0.90 &  0.37 &  80.4 &  11.5 &   96.1 \\ 
  2.19 & -11.77 &  1.16 &  0.50 &  96.9 &  12.4 &   67.3 \\ 
 -2.85 & -11.73 &  3.56 &  0.68 &   3.1 &   5.3 &   75.7 \\ 
 -2.56 & -11.73 &  2.67 &  0.56 & 165.7 &   6.0 &   88.8 \\ 
 -2.26 & -11.73 &  2.19 &  0.38 & 165.2 &   4.9 &  107.2 \\ 
 -1.96 & -11.73 &  0.92 &  0.25 & 177.8 &   7.9 &  154.3 \\ 
 -1.67 & -11.73 &  0.72 &  0.17 & 169.6 &   6.6 &  232.7 \\ 
 -1.37 & -11.73 &  0.35 &  0.13 & 175.6 &  11.7 &  342.5 \\ 
 -1.07 & -11.73 &  0.32 &  0.10 &  45.1 &   9.5 &  506.3 \\ 
 -0.78 & -11.73 &  1.02 &  0.08 &  45.0 &   2.4 &  524.8 \\ 
 -0.48 & -11.73 &  1.70 &  0.16 &  43.7 &   3.3 &  325.7 \\ 
 -0.18 & -11.73 &  1.76 &  0.15 &  43.8 &   2.4 &  234.9 \\ 
  0.11 & -11.73 &  1.99 &  0.21 &  32.0 &   3.0 &  189.6 \\ 
  0.41 & -11.73 &  1.99 &  0.30 &  42.3 &   4.3 &  141.8 \\ 
  0.71 & -11.73 &  1.46 &  0.46 &  41.4 &   9.1 &  120.0 \\ 
 -1.39 & -11.48 &  0.61 &  0.30 &  86.7 &  14.5 &  319.4 \\ 
 -1.09 & -11.48 &  0.42 &  0.21 &  84.5 &  14.5 &  422.2 \\ 
 -0.80 & -11.48 &  1.22 &  0.21 &  60.9 &   4.9 &  432.3 \\ 
  0.71 & -11.47 &  2.97 &  0.90 &  34.2 &   8.9 &   74.2 \\ 
  1.00 & -11.47 &  0.90 &  0.45 &  46.1 &  14.3 &   73.1 \\ 
  1.30 & -11.47 &  1.54 &  0.44 &  51.5 &   8.2 &   80.8 \\ 
  1.89 & -11.47 &  1.51 &  0.29 & 110.4 &   5.6 &  118.6 \\ 
 -2.85 & -11.43 &  3.00 &  1.04 & 175.9 &   6.6 &   83.4 \\ 
 -2.56 & -11.43 &  1.13 &  0.44 & 176.8 &  11.2 &   98.1 \\ 
 -2.26 & -11.43 &  2.15 &  0.44 & 169.5 &   4.6 &  131.4 \\ 
 -1.96 & -11.43 &  0.89 &  0.20 & 173.2 &   9.4 &  182.0 \\ 
 -1.67 & -11.43 &  1.16 &  0.12 &  11.3 &   3.1 &  257.0 \\ 
 -1.37 & -11.43 &  0.80 &  0.12 &   9.8 &   4.5 &  411.1 \\ 
 -1.07 & -11.43 &  0.97 &  0.07 &  23.7 &   2.2 &  886.1 \\ 
 -0.78 & -11.43 &  1.13 &  0.11 &  29.7 &   1.9 &  952.8 \\ 
 -0.48 & -11.43 &  1.67 &  0.14 &  37.3 &   2.0 &  466.0 \\ 
 -0.18 & -11.43 &  2.18 &  0.18 &  38.9 &   1.7 &  285.6 \\ 
  0.11 & -11.43 &  2.32 &  0.17 &  39.6 &   2.2 &  213.3 \\ 
  0.41 & -11.43 &  1.88 &  0.25 &  33.6 &   3.7 &  159.0 \\ 
  0.71 & -11.43 &  1.17 &  0.45 &  33.9 &  10.9 &  127.5 \\ 
  1.00 & -11.43 &  1.65 &  0.68 &  46.5 &  11.8 &  127.7 \\ 
  1.89 & -11.18 &  1.26 &  0.47 &  64.0 &  10.7 &   89.9 \\ 
 -2.85 & -11.14 &  3.14 &  0.94 & 175.5 &   8.6 &   93.6 \\ 
 -2.56 & -11.14 &  3.00 &  0.85 & 166.6 &   5.8 &  107.8 \\ 
 -2.26 & -11.14 &  2.68 &  0.33 & 173.9 &   3.5 &  153.1 \\ 
 -1.96 & -11.14 &  1.58 &  0.21 & 177.5 &   3.9 &  220.5 \\ 
 -1.67 & -11.14 &  1.17 &  0.15 &  13.3 &   3.9 &  316.0 \\ 
 -1.37 & -11.14 &  1.23 &  0.24 &  20.8 &   5.1 &  442.0 \\ 
 -1.07 & -11.14 &  1.76 &  0.07 &  24.2 &   1.1 &  766.0 \\ 
 -0.78 & -11.14 &  2.04 &  0.11 &  24.0 &   1.4 &  947.8 \\ 
 -0.48 & -11.14 &  2.57 &  0.10 &  27.7 &   1.0 &  491.0 \\ 
 -0.18 & -11.14 &  2.78 &  0.12 &  29.2 &   1.3 &  308.0 \\ 
  0.11 & -11.14 &  2.25 &  0.19 &  26.8 &   2.5 &  222.6 \\ 
  0.41 & -11.14 &  2.40 &  0.28 &  29.4 &   3.3 &  157.8 \\ 
  0.71 & -11.14 &  1.54 &  0.45 &  38.1 &   8.8 &  118.4 \\ 
  1.00 & -11.14 &  1.49 &  0.56 &  32.7 &  10.9 &  133.8 \\ 
 -2.56 & -10.84 &  2.97 &  0.63 &   0.2 &   6.1 &  135.7 \\ 
 -2.26 & -10.84 &  1.43 &  0.39 &   5.0 &   7.8 &  190.1 \\ 
 -1.96 & -10.84 &  2.01 &  0.27 &  22.5 &   3.9 &  270.5 \\ 
 -1.67 & -10.84 &  2.13 &  0.23 &  14.3 &   3.1 &  351.8 \\ 
 -1.37 & -10.84 &  1.61 &  0.25 &  25.2 &   4.9 &  477.7 \\ 
 -1.07 & -10.84 &  2.64 &  0.17 &  24.3 &   1.7 &  657.5 \\ 
 -0.78 & -10.84 &  2.82 &  0.10 &  25.5 &   1.0 &  709.0 \\ 
 -0.48 & -10.84 &  2.71 &  0.10 &  26.8 &   1.1 &  450.6 \\ 
 -0.18 & -10.84 &  2.96 &  0.12 &  25.5 &   1.2 &  349.8 \\ 
  0.11 & -10.84 &  2.44 &  0.17 &  29.2 &   2.0 &  232.6 \\ 
  0.41 & -10.84 &  2.25 &  0.30 &  20.1 &   3.8 &  160.1 \\ 
  0.71 & -10.84 &  2.63 &  0.50 &  15.2 &   5.4 &  120.8 \\ 
  1.00 & -10.84 &  2.52 &  0.70 &  13.5 &   8.2 &  126.5 \\ 
 -1.05 & -10.60 &  1.51 &  0.15 &  37.6 &   2.9 &  832.2 \\ 
 -0.75 & -10.60 &  1.91 &  0.17 &  17.9 &   2.5 &  988.7 \\ 
 -0.45 & -10.60 &  2.47 &  0.37 &  26.2 &   4.3 &  400.0 \\ 
 -0.16 & -10.60 &  2.13 &  0.35 &  18.5 &   4.7 &  276.3 \\ 
 -2.56 & -10.54 &  2.13 &  0.70 &  20.2 &   9.5 &  160.8 \\ 
 -2.26 & -10.54 &  2.26 &  0.58 &   8.6 &   7.4 &  194.2 \\ 
 -1.96 & -10.54 &  1.60 &  0.42 &  12.7 &   7.5 &  280.4 \\ 
 -1.67 & -10.54 &  1.90 &  0.38 &  32.6 &   5.7 &  444.7 \\ 
 -1.07 & -10.54 &  3.44 &  0.31 &  21.5 &   2.6 &  547.2 \\ 
 -0.78 & -10.54 &  3.54 &  0.22 &  22.8 &   1.8 &  573.2 \\ 
 -0.48 & -10.54 &  3.24 &  0.17 &  26.6 &   1.5 &  449.9 \\ 
 -0.18 & -10.54 &  3.05 &  0.19 &  25.3 &   1.7 &  368.8 \\ 
  0.11 & -10.54 &  2.84 &  0.21 &  28.2 &   2.1 &  273.7 \\ 
  0.41 & -10.54 &  3.07 &  0.39 &  21.3 &   3.6 &  183.4 \\ 
  0.71 & -10.54 &  2.38 &  0.61 &  26.2 &   7.3 &  198.2 \\ 
  1.30 & -10.54 &  2.74 &  1.13 & 170.5 &  11.1 &  125.4 \\ 
 -2.53 & -10.31 &  2.32 &  0.58 &  10.0 &   7.1 &  132.6 \\ 
 -2.23 & -10.31 &  1.91 &  0.48 &  20.3 &   7.1 &  169.8 \\ 
 -1.94 & -10.31 &  2.12 &  0.33 &  18.2 &   4.4 &  232.9 \\ 
 -1.64 & -10.31 &  2.79 &  0.25 &  19.5 &   2.5 &  291.5 \\ 
 -1.34 & -10.31 &  1.89 &  0.08 &  24.1 &   1.2 &  531.5 \\ 
 -1.05 & -10.31 &  3.07 &  0.08 &  27.3 &   0.9 &  665.4 \\ 
 -0.75 & -10.31 &  3.33 &  0.07 &  26.4 &   0.6 &  647.1 \\ 
 -0.45 & -10.31 &  3.02 &  0.05 &  26.4 &   0.5 &  460.2 \\ 
 -0.16 & -10.31 &  2.50 &  0.09 &  28.0 &   1.1 &  341.8 \\ 
  0.14 & -10.31 &  2.12 &  0.47 &  21.9 &   6.3 &  236.3 \\ 
  0.11 & -10.25 &  2.20 &  0.37 &  27.9 &   4.9 &  331.0 \\ 
  0.41 & -10.25 &  3.12 &  0.79 &  32.6 &   7.3 &  245.8 \\ 
  0.71 & -10.25 &  2.51 &  0.66 &  38.7 &   7.6 &  220.7 \\ 
  1.00 & -10.25 &  1.81 &  0.89 &  33.0 &  14.7 &  165.2 \\ 
 -2.83 & -10.01 &  2.12 &  0.52 &  10.6 &   7.0 &  141.9 \\ 
 -2.53 & -10.01 &  2.22 &  0.30 &  14.8 &   3.8 &  178.3 \\ 
 -2.23 & -10.01 &  2.11 &  0.25 &  21.9 &   3.4 &  200.8 \\ 
 -1.94 & -10.01 &  2.62 &  0.21 &  28.6 &   2.4 &  249.9 \\ 
 -1.64 & -10.01 &  2.54 &  0.07 &  22.0 &   0.8 &  406.7 \\ 
 -1.34 & -10.01 &  2.57 &  0.04 &  27.6 &   0.5 &  531.7 \\ 
 -1.05 & -10.01 &  2.98 &  0.05 &  27.7 &   0.5 &  788.2 \\ 
 -0.75 & -10.01 &  3.45 &  0.07 &  26.2 &   0.6 &  708.0 \\ 
 -0.45 & -10.01 &  2.97 &  0.06 &  27.6 &   0.5 &  535.2 \\ 
 -0.16 & -10.01 &  2.60 &  0.05 &  23.8 &   0.5 &  377.8 \\ 
  0.14 & -10.01 &  2.37 &  0.14 &  23.1 &   1.7 &  279.3 \\ 
 -2.83 &  -9.71 &  2.55 &  0.34 &  24.1 &   3.8 &  154.8 \\ 
 -2.53 &  -9.71 &  2.52 &  0.30 &  23.4 &   3.4 &  174.1 \\ 
 -2.23 &  -9.71 &  2.70 &  0.26 &  23.3 &   2.7 &  196.7 \\ 
 -1.94 &  -9.71 &  2.67 &  0.18 &  28.8 &   1.9 &  226.8 \\ 
 -1.64 &  -9.71 &  2.76 &  0.07 &  28.2 &   0.7 &  358.5 \\ 
 -1.34 &  -9.71 &  2.77 &  0.06 &  29.1 &   0.5 &  538.5 \\ 
 -1.05 &  -9.71 &  2.50 &  0.06 &  26.8 &   0.7 &  999.4 \\ 
 -0.75 &  -9.71 &  2.01 &  0.04 &  31.0 &   0.6 & 1257.8 \\ 
 -0.45 &  -9.71 &  2.41 &  0.07 &  26.3 &   0.8 &  697.9 \\ 
 -0.16 &  -9.71 &  2.53 &  0.04 &  22.2 &   0.5 &  423.3 \\ 
  0.14 &  -9.71 &  2.47 &  0.08 &  23.0 &   0.9 &  296.5 \\ 
  0.44 &  -9.71 &  2.67 &  0.26 &  20.0 &   2.8 &  215.7 \\ 
  0.73 &  -9.71 &  2.87 &  0.37 &  10.1 &   3.5 &  164.1 \\ 
  1.03 &  -9.71 &  2.60 &  0.46 &  11.4 &   5.0 &  102.8 \\ 
  1.33 &  -9.71 &  2.76 &  0.55 &   6.6 &   5.4 &   82.3 \\ 
 -1.94 &  -9.42 &  3.00 &  0.18 &  32.7 &   1.7 &  237.0 \\ 
  0.44 &  -9.42 &  2.51 &  0.18 &  17.7 &   2.1 &  235.6 \\ 
 -0.75 &  -9.42 &  1.06 &  0.04 &  33.3 &   1.3 & 2100.0 \\ 
 -2.83 &  -9.42 &  2.30 &  0.47 &  42.1 &   5.8 &  118.4 \\ 
 -2.53 &  -9.42 &  2.97 &  0.36 &  36.7 &   3.5 &  138.9 \\ 
 -2.23 &  -9.42 &  2.26 &  0.35 &  32.7 &   4.5 &  159.6 \\ 
 -1.64 &  -9.42 &  2.91 &  0.05 &  32.9 &   0.5 &  399.5 \\ 
 -1.34 &  -9.42 &  2.88 &  0.04 &  32.4 &   0.5 &  531.7 \\ 
 -1.05 &  -9.42 &  2.10 &  0.06 &  28.5 &   0.6 & 1011.5 \\ 
 -0.45 &  -9.42 &  1.89 &  0.04 &  24.5 &   0.6 &  952.9 \\ 
 -0.16 &  -9.42 &  2.67 &  0.06 &  19.4 &   0.6 &  473.5 \\ 
  0.14 &  -9.42 &  2.46 &  0.12 &  17.5 &   1.1 &  316.0 \\ 
  0.73 &  -9.42 &  2.81 &  0.31 &   8.6 &   3.1 &  158.7 \\ 
  1.03 &  -9.42 &  2.80 &  0.37 &   3.5 &   3.7 &  111.1 \\ 
  1.33 &  -9.42 &  2.66 &  0.49 &   2.1 &   5.2 &   94.6 \\ 
  1.62 &  -9.42 &  3.27 &  0.54 & 176.7 &   4.8 &   81.9 \\ 
 -2.83 &  -9.12 &  3.20 &  0.57 &  44.3 &   5.1 &  106.2 \\ 
 -2.53 &  -9.12 &  3.22 &  0.42 &  33.2 &   3.7 &  131.5 \\ 
 -2.23 &  -9.12 &  2.85 &  0.30 &  34.9 &   2.9 &  172.5 \\ 
 -1.94 &  -9.12 &  2.78 &  0.17 &  34.8 &   1.8 &  267.1 \\ 
 -1.64 &  -9.12 &  2.74 &  0.05 &  37.5 &   0.9 &  387.6 \\ 
 -1.34 &  -9.12 &  2.38 &  0.05 &  37.0 &   0.8 &  516.0 \\ 
 -1.05 &  -9.12 &  2.02 &  0.04 &  35.1 &   0.5 &  807.4 \\ 
 -0.75 &  -9.12 &  1.49 &  0.05 &  34.0 &   1.1 & 1238.4 \\ 
 -0.45 &  -9.12 &  2.33 &  0.05 &  22.6 &   0.6 & 1134.4 \\ 
 -0.16 &  -9.12 &  2.88 &  0.05 &  14.1 &   0.4 &  508.9 \\ 
  0.14 &  -9.12 &  2.52 &  0.07 &  13.1 &   0.9 &  311.7 \\ 
  0.44 &  -9.12 &  2.34 &  0.22 &   8.3 &   4.7 &  231.2 \\ 
  0.73 &  -9.12 &  2.77 &  0.29 & 178.7 &   3.0 &  156.0 \\ 
  1.03 &  -9.12 &  3.05 &  0.38 &   2.0 &   3.6 &  127.1 \\ 
  1.33 &  -9.12 &  2.45 &  0.44 &   7.6 &   4.9 &  103.7 \\ 
  1.62 &  -9.12 &  2.58 &  0.56 &   2.5 &   6.1 &   82.1 \\ 
 -2.83 &  -8.82 &  3.02 &  0.82 &  35.1 &   7.8 &  103.3 \\ 
 -2.53 &  -8.82 &  3.58 &  0.40 &  39.7 &   3.2 &  144.0 \\ 
 -2.23 &  -8.82 &  3.41 &  0.29 &  37.4 &   2.5 &  174.9 \\ 
 -1.94 &  -8.82 &  2.96 &  0.19 &  42.3 &   1.9 &  236.1 \\ 
 -1.64 &  -8.82 &  2.80 &  0.11 &  39.6 &   0.9 &  342.1 \\ 
 -1.34 &  -8.82 &  2.28 &  0.04 &  40.1 &   0.6 &  455.7 \\ 
 -1.05 &  -8.82 &  1.94 &  0.03 &  39.0 &   0.7 &  615.0 \\ 
 -0.75 &  -8.82 &  2.07 &  0.05 &  38.3 &   0.6 &  725.1 \\ 
 -0.45 &  -8.82 &  1.79 &  0.04 &  29.1 &   0.8 &  949.0 \\ 
 -0.16 &  -8.82 &  2.20 &  0.05 &  13.0 &   0.6 &  552.6 \\ 
  0.14 &  -8.82 &  2.65 &  0.07 &  10.5 &   0.8 &  316.5 \\ 
  0.44 &  -8.82 &  2.37 &  0.17 &   4.3 &   2.1 &  243.6 \\ 
  0.73 &  -8.82 &  4.03 &  0.59 &   4.6 &   4.2 &  184.9 \\ 
  1.03 &  -8.82 &  3.52 &  0.40 &   1.6 &   3.2 &  151.1 \\ 
  1.33 &  -8.82 &  3.59 &  0.47 &   3.1 &   3.6 &  115.2 \\ 
  1.62 &  -8.82 &  5.65 &  0.70 &   1.1 &   3.5 &   83.8 \\ 
 -2.53 &  -8.53 &  4.10 &  0.60 &  39.9 &   4.1 &  157.4 \\ 
 -2.23 &  -8.53 &  3.34 &  0.60 &  35.3 &   5.2 &  148.4 \\ 
 -1.94 &  -8.53 &  3.62 &  0.32 &  43.7 &   2.6 &  193.3 \\ 
 -1.64 &  -8.53 &  3.37 &  0.28 &  37.8 &   2.4 &  265.5 \\ 
 -1.34 &  -8.53 &  2.32 &  0.09 &  45.1 &   1.1 &  381.8 \\ 
 -1.05 &  -8.53 &  2.19 &  0.06 &  41.5 &   0.7 &  467.0 \\ 
 -0.75 &  -8.53 &  2.18 &  0.07 &  36.3 &   0.6 &  508.9 \\ 
 -0.45 &  -8.53 &  1.86 &  0.04 &  28.2 &   0.8 &  566.1 \\ 
 -0.16 &  -8.53 &  1.76 &  0.06 &  15.4 &   1.0 &  482.0 \\ 
  0.14 &  -8.53 &  1.96 &  0.10 &  10.6 &   1.9 &  437.9 \\ 
  0.44 &  -8.53 &  2.34 &  0.18 & 178.7 &   3.1 &  268.5 \\ 
  0.73 &  -8.53 &  2.52 &  0.28 & 178.0 &   3.2 &  194.9 \\ 
  1.03 &  -8.53 &  3.01 &  0.33 & 179.2 &   3.2 &  152.6 \\ 
  1.33 &  -8.53 &  4.30 &  0.69 & 176.5 &   4.7 &  111.4 \\ 
  1.62 &  -8.53 &  4.61 &  0.71 & 177.4 &   4.4 &   80.0 \\ 
 -2.62 &  -8.41 &  2.39 &  0.82 &  57.7 &   9.9 &  109.7 \\ 
 -2.32 &  -8.41 &  3.16 &  0.81 &  48.7 &   7.2 &  119.1 \\ 
 -2.03 &  -8.41 &  5.66 &  0.63 &  50.0 &   3.1 &  146.1 \\ 
 -1.73 &  -8.41 &  4.43 &  0.47 &  50.1 &   3.0 &  187.9 \\ 
 -0.84 &  -8.40 &  2.20 &  0.22 &  39.4 &   2.9 &  408.3 \\ 
 -0.55 &  -8.40 &  1.93 &  0.14 &  33.9 &   2.1 &  489.2 \\ 
 -0.25 &  -8.40 &  1.57 &  0.13 &  21.3 &   2.4 &  542.8 \\ 
  0.05 &  -8.40 &  1.47 &  0.14 &  17.6 &   2.7 &  496.2 \\ 
  0.34 &  -8.40 &  1.68 &  0.22 &   5.0 &   3.8 &  411.2 \\ 
  0.64 &  -8.40 &  3.17 &  0.28 & 177.1 &   2.6 &  212.4 \\ 
  0.94 &  -8.40 &  4.04 &  0.49 & 175.7 &   3.5 &  155.4 \\ 
  1.23 &  -8.40 &  3.35 &  0.55 &   5.2 &   4.8 &  102.9 \\ 
  1.53 &  -8.40 &  5.43 &  0.86 &   1.6 &   4.5 &   67.0 \\ 
 -1.64 &  -8.23 &  3.82 &  0.79 &  45.9 &   5.9 &  191.9 \\ 
 -1.34 &  -8.23 &  3.41 &  0.66 &  46.3 &   5.5 &  217.2 \\ 
 -1.05 &  -8.23 &  2.36 &  0.21 &  43.6 &   2.6 &  320.3 \\ 
 -0.75 &  -8.23 &  2.01 &  0.15 &  41.1 &   2.0 &  399.4 \\ 
 -0.45 &  -8.23 &  1.61 &  0.13 &  28.0 &   2.2 &  541.4 \\ 
 -0.16 &  -8.23 &  1.87 &  0.15 &  19.6 &   2.4 &  470.5 \\ 
  0.14 &  -8.23 &  1.60 &  0.23 &  23.3 &   4.2 &  415.8 \\ 
  0.44 &  -8.23 &  1.63 &  0.26 & 174.5 &   4.7 &  286.0 \\ 
  0.73 &  -8.23 &  3.25 &  0.35 & 175.5 &   3.1 &  183.3 \\ 
  1.03 &  -8.23 &  3.35 &  0.45 & 178.0 &   3.9 &  134.7 \\ 
  1.33 &  -8.23 &  3.04 &  0.50 & 179.1 &   4.7 &  103.6 \\ 
 -2.92 &  -8.11 &  6.01 &  1.51 &  59.9 &   7.0 &   73.1 \\ 
 -2.62 &  -8.11 &  3.18 &  0.95 &  50.5 &   8.2 &   94.0 \\ 
 -2.32 &  -8.11 &  2.78 &  0.73 &  53.1 &   7.3 &  103.8 \\ 
 -2.03 &  -8.11 &  4.59 &  0.59 &  48.3 &   3.6 &  138.3 \\ 
 -1.73 &  -8.11 &  3.86 &  0.45 &  49.8 &   3.3 &  181.1 \\ 
 -1.43 &  -8.11 &  3.10 &  0.46 &  44.5 &   4.3 &  185.2 \\ 
 -1.14 &  -8.10 &  2.45 &  0.35 &  47.4 &   4.0 &  293.4 \\ 
 -0.84 &  -8.10 &  2.06 &  0.11 &  40.8 &   1.5 &  350.5 \\ 
 -0.55 &  -8.10 &  1.89 &  0.07 &  35.3 &   1.0 &  488.3 \\ 
 -0.25 &  -8.10 &  1.68 &  0.07 &  20.8 &   1.2 &  483.9 \\ 
  0.05 &  -8.10 &  1.42 &  0.09 &  13.7 &   1.8 &  341.5 \\ 
  0.34 &  -8.10 &  1.29 &  0.32 &   5.3 &   2.8 &  337.9 \\ 
  0.64 &  -8.10 &  1.71 &  0.22 & 179.8 &   4.9 &  257.1 \\ 
  0.94 &  -8.10 &  3.21 &  0.38 & 176.6 &   3.4 &  152.3 \\ 
  1.23 &  -8.10 &  4.29 &  0.60 & 176.3 &   4.0 &  103.1 \\ 
  1.53 &  -8.10 &  6.54 &  0.97 & 178.7 &   4.3 &   75.1 \\ 
  1.83 &  -8.10 &  5.72 &  1.34 &   5.0 &   6.4 &   53.4 \\ 
 -2.92 &  -7.81 &  5.35 &  1.69 &  70.3 &   9.0 &   59.8 \\ 
 -2.62 &  -7.81 &  3.94 &  1.11 &  44.2 &   8.1 &   71.4 \\ 
 -2.32 &  -7.81 &  2.66 &  1.10 &  52.2 &  11.5 &   92.0 \\ 
 -2.03 &  -7.81 &  3.89 &  0.63 &  49.6 &   4.5 &  126.0 \\ 
 -1.73 &  -7.81 &  2.47 &  0.48 &  46.2 &   5.5 &  161.3 \\ 
 -1.43 &  -7.81 &  3.99 &  0.48 &  48.5 &   3.4 &  168.6 \\ 
 -1.14 &  -7.81 &  2.63 &  0.17 &  48.9 &   1.8 &  255.4 \\ 
 -0.84 &  -7.81 &  1.96 &  0.10 &  47.3 &   1.4 &  286.9 \\ 
 -0.55 &  -7.81 &  1.58 &  0.07 &  41.1 &   1.3 &  393.6 \\ 
 -0.25 &  -7.81 &  1.22 &  0.05 &  20.4 &   1.2 &  493.1 \\ 
  0.05 &  -7.81 &  1.28 &  0.08 &  18.1 &   1.7 &  357.1 \\ 
  0.34 &  -7.81 &  1.23 &  0.20 &  14.9 &   3.1 &  293.6 \\ 
  0.64 &  -7.81 &  1.69 &  0.16 &   7.6 &   3.3 &  205.4 \\ 
  0.94 &  -7.81 &  2.61 &  0.31 &   2.4 &   3.4 &  145.8 \\ 
  1.23 &  -7.81 &  3.18 &  0.42 &   6.8 &   3.7 &  102.1 \\ 
  1.53 &  -7.81 &  4.53 &  0.57 &   9.3 &   3.4 &   79.9 \\ 
  1.83 &  -7.81 &  6.60 &  1.03 &  15.7 &   4.4 &   63.1 \\ 
  2.12 &  -7.81 &  5.48 &  2.34 & 175.8 &  12.2 &   61.4 \\ 
 -2.92 &  -7.52 &  5.87 &  1.87 &  51.4 &   8.7 &   48.0 \\ 
 -2.62 &  -7.52 &  3.65 &  1.60 &  66.3 &  12.6 &   59.8 \\ 
 -2.03 &  -7.52 &  4.31 &  0.66 &  53.6 &   4.2 &  126.8 \\ 
 -1.73 &  -7.52 &  3.89 &  0.58 &  52.3 &   4.1 &  146.0 \\ 
 -1.43 &  -7.52 &  3.23 &  0.60 &  68.5 &   5.3 &  147.5 \\ 
 -2.32 &  -7.52 &  3.48 &  1.02 &  54.6 &   8.2 &   82.8 \\ 
 -1.14 &  -7.51 &  2.06 &  0.19 &  55.3 &   2.6 &  219.2 \\ 
 -0.84 &  -7.51 &  1.64 &  0.12 &  50.4 &   2.0 &  260.3 \\ 
 -0.55 &  -7.51 &  1.36 &  0.07 &  41.5 &   1.4 &  420.6 \\ 
 -0.25 &  -7.51 &  0.68 &  0.06 &  31.5 &   2.4 &  483.3 \\ 
  0.05 &  -7.51 &  0.79 &  0.07 &  23.5 &   2.6 &  341.3 \\ 
  0.34 &  -7.51 &  1.36 &  0.10 &  14.9 &   2.9 &  288.4 \\ 
  0.64 &  -7.51 &  1.77 &  0.15 &  13.0 &   2.4 &  205.4 \\ 
  0.94 &  -7.51 &  3.02 &  0.43 &  17.3 &   4.1 &  131.0 \\ 
  1.23 &  -7.51 &  3.72 &  0.58 &  11.8 &   3.2 &  107.0 \\ 
  1.53 &  -7.51 &  5.74 &  0.50 &  14.8 &   2.4 &   89.1 \\ 
  1.83 &  -7.51 &  6.87 &  0.60 &  12.8 &   2.4 &   78.9 \\ 
  2.12 &  -7.51 &  7.59 &  1.49 &  18.2 &   5.6 &   80.4 \\ 
  2.42 &  -7.51 &  7.87 &  2.05 &  19.1 &   7.5 &   79.0 \\ 
 -2.92 &  -7.22 &  8.08 &  3.53 &  64.2 &  12.5 &   40.0 \\ 
 -2.62 &  -7.22 &  2.88 &  1.15 &  50.2 &  11.4 &   54.3 \\ 
 -2.32 &  -7.22 &  3.67 &  0.81 &  53.2 &   6.4 &   80.7 \\ 
 -2.03 &  -7.22 &  2.55 &  0.40 &  47.5 &   4.5 &  120.9 \\ 
 -1.73 &  -7.22 &  3.57 &  0.63 &  57.0 &   4.7 &  112.6 \\ 
 -1.43 &  -7.22 &  4.31 &  1.12 &  62.1 &   7.4 &  126.9 \\ 
 -1.14 &  -7.21 &  2.41 &  0.24 &  55.1 &   2.8 &  183.7 \\ 
 -0.84 &  -7.21 &  1.79 &  0.11 &  50.1 &   1.7 &  228.6 \\ 
 -0.55 &  -7.21 &  1.74 &  0.07 &  35.8 &   1.1 &  342.3 \\ 
 -0.25 &  -7.21 &  1.12 &  0.06 &  36.4 &   1.6 &  346.8 \\ 
  0.05 &  -7.21 &  0.94 &  0.08 &  42.4 &   2.4 &  261.7 \\ 
  0.34 &  -7.21 &  1.45 &  0.10 &  27.6 &   2.1 &  241.8 \\ 
  0.64 &  -7.21 &  2.34 &  0.17 &  20.9 &   2.0 &  192.9 \\ 
  0.94 &  -7.21 &  3.85 &  0.32 &  20.4 &   2.4 &  141.9 \\ 
  1.23 &  -7.21 &  4.20 &  0.43 &  20.2 &   2.9 &  114.7 \\ 
  1.53 &  -7.21 &  5.55 &  0.56 &  17.4 &   2.9 &  107.3 \\ 
  1.83 &  -7.21 &  5.25 &  0.44 &  13.4 &   2.4 &  105.4 \\ 
  2.12 &  -7.21 &  6.23 &  1.04 &  18.5 &   4.8 &  104.4 \\ 
  2.42 &  -7.21 &  5.79 &  1.44 &   6.2 &   7.0 &   75.9 \\ 
 -2.62 &  -6.92 &  3.12 &  1.19 &  48.5 &  11.0 &   50.5 \\ 
 -2.32 &  -6.92 &  3.63 &  0.79 &  47.2 &   5.6 &   77.9 \\ 
 -2.03 &  -6.92 &  3.94 &  0.46 &  57.0 &   3.4 &  104.5 \\ 
 -1.73 &  -6.92 &  3.57 &  0.58 &  64.3 &   4.6 &   99.0 \\ 
 -1.43 &  -6.92 &  2.64 &  0.77 &  46.8 &   8.5 &  107.3 \\ 
 -0.84 &  -6.92 &  2.31 &  0.11 &  51.7 &   1.3 &  205.3 \\ 
 -0.55 &  -6.92 &  1.86 &  0.07 &  41.8 &   1.1 &  302.4 \\ 
  0.05 &  -6.92 &  1.47 &  0.09 &  37.7 &   1.7 &  227.0 \\ 
  0.34 &  -6.92 &  2.24 &  0.10 &  32.8 &   1.2 &  208.2 \\ 
  0.64 &  -6.92 &  3.15 &  0.16 &  31.0 &   1.3 &  189.4 \\ 
  0.94 &  -6.92 &  4.26 &  0.33 &  34.7 &   3.1 &  150.6 \\ 
  1.23 &  -6.92 &  4.89 &  0.37 &  21.4 &   2.2 &  116.7 \\ 
  1.53 &  -6.92 &  5.75 &  0.48 &  17.5 &   2.4 &  104.2 \\ 
  1.83 &  -6.92 &  6.80 &  0.71 &  21.5 &   3.0 &   95.6 \\ 
  2.12 &  -6.92 &  7.27 &  1.26 &  13.7 &   4.9 &   82.4 \\ 
  2.42 &  -6.92 &  9.37 &  1.73 &   8.5 &   5.1 &   64.9 \\ 
 -0.25 &  -6.92 &  1.28 &  0.07 &  37.0 &   1.5 &  304.8 \\ 
 -1.14 &  -6.92 &  2.84 &  0.19 &  54.6 &   1.9 &  157.0 \\ 
 -2.62 &  -6.63 &  3.92 &  1.39 &  41.6 &   9.9 &   46.0 \\ 
 -2.32 &  -6.63 &  4.33 &  0.93 &  50.2 &   6.2 &   65.5 \\ 
 -2.03 &  -6.63 &  3.43 &  0.72 &  57.8 &   5.9 &   77.5 \\ 
 -1.73 &  -6.63 &  2.88 &  0.73 &  44.3 &   7.3 &   86.3 \\ 
 -1.43 &  -6.63 &  4.21 &  0.84 &  53.1 &   5.6 &   94.2 \\ 
 -1.14 &  -6.62 &  3.71 &  0.28 &  48.9 &   2.2 &  150.3 \\ 
 -0.84 &  -6.62 &  2.33 &  0.14 &  53.3 &   1.7 &  194.2 \\ 
 -0.55 &  -6.62 &  1.83 &  0.08 &  44.5 &   1.3 &  268.5 \\ 
 -0.25 &  -6.62 &  1.39 &  0.06 &  41.6 &   1.2 &  342.2 \\ 
  0.05 &  -6.62 &  1.96 &  0.08 &  39.0 &   1.2 &  235.7 \\ 
  0.34 &  -6.62 &  2.64 &  0.10 &  33.0 &   1.1 &  200.9 \\ 
  0.64 &  -6.62 &  3.64 &  0.24 &  33.8 &   1.5 &  172.1 \\ 
  0.94 &  -6.62 &  4.56 &  0.28 &  30.2 &   1.8 &  154.8 \\ 
  1.23 &  -6.62 &  5.93 &  0.55 &  23.8 &   2.7 &  115.1 \\ 
  1.53 &  -6.62 &  5.13 &  0.94 &  22.3 &   5.2 &   96.5 \\ 
  1.83 &  -6.62 &  4.93 &  0.90 &  13.5 &   6.1 &   75.2 \\ 
  2.12 &  -6.62 &  5.48 &  1.78 &   3.4 &   9.2 &   57.2 \\ 
  2.42 &  -6.62 &  5.69 &  2.38 &  11.9 &  11.6 &   45.5 \\ 
 -2.62 &  -6.33 &  4.39 &  1.81 &  49.9 &  11.7 &   43.3 \\ 
 -2.32 &  -6.33 &  2.73 &  1.05 &  61.2 &  11.0 &   55.9 \\ 
 -2.03 &  -6.33 &  3.98 &  0.92 &  44.9 &   6.6 &   66.7 \\ 
 -1.73 &  -6.33 &  4.16 &  0.83 &  51.7 &   5.6 &   78.1 \\ 
 -1.43 &  -6.33 &  3.78 &  1.09 &  52.5 &   8.2 &   92.4 \\ 
 -1.14 &  -6.32 &  3.40 &  0.31 &  55.8 &   2.5 &  148.0 \\ 
 -0.84 &  -6.32 &  2.45 &  0.16 &  49.8 &   1.8 &  200.2 \\ 
 -0.55 &  -6.32 &  1.56 &  0.09 &  48.8 &   1.7 &  278.9 \\ 
 -0.25 &  -6.32 &  1.62 &  0.06 &  45.0 &   1.1 &  395.0 \\ 
  0.05 &  -6.32 &  1.88 &  0.07 &  41.6 &   1.0 &  311.9 \\ 
  0.34 &  -6.32 &  2.98 &  0.13 &  35.3 &   1.2 &  199.5 \\ 
  0.64 &  -6.32 &  3.87 &  0.18 &  32.0 &   1.3 &  157.3 \\ 
  0.94 &  -6.32 &  4.80 &  0.26 &  33.5 &   1.5 &  138.5 \\ 
  1.23 &  -6.32 &  4.19 &  0.48 &  24.7 &   3.2 &  106.1 \\ 
  1.53 &  -6.32 &  3.84 &  0.73 &  16.6 &   5.3 &   80.7 \\ 
  1.83 &  -6.32 &  7.03 &  1.42 &  13.5 &   5.7 &   65.5 \\ 
 -2.62 &  -6.03 &  4.82 &  2.40 &  50.1 &  14.7 &   37.6 \\ 
 -2.32 &  -6.03 &  4.06 &  1.35 &  54.3 &   9.6 &   49.4 \\ 
 -2.03 &  -6.03 &  2.83 &  0.84 &  57.3 &   8.7 &   64.8 \\ 
 -1.73 &  -6.03 &  3.57 &  0.81 &  62.9 &   6.5 &   73.4 \\ 
 -1.43 &  -6.03 &  3.60 &  0.86 &  53.4 &   6.6 &   80.6 \\ 
 -1.14 &  -6.03 &  4.07 &  0.29 &  55.4 &   2.0 &  143.0 \\ 
 -0.84 &  -6.03 &  1.91 &  0.14 &  58.2 &   2.1 &  222.4 \\ 
 -0.55 &  -6.03 &  1.88 &  0.11 &  48.3 &   1.6 &  245.9 \\ 
 -0.25 &  -6.03 &  2.06 &  0.08 &  44.9 &   1.1 &  291.9 \\ 
  0.05 &  -6.03 &  1.83 &  0.07 &  44.3 &   1.1 &  339.1 \\ 
  0.34 &  -6.03 &  2.28 &  0.10 &  34.4 &   1.2 &  227.4 \\ 
  0.64 &  -6.03 &  3.31 &  0.18 &  36.4 &   1.5 &  160.6 \\ 
  0.94 &  -6.03 &  3.67 &  0.40 &  31.4 &   2.5 &  109.7 \\ 
  1.23 &  -6.03 &  3.30 &  0.61 &  34.4 &   5.5 &   84.8 \\ 
 -2.32 &  -5.74 &  4.46 &  1.45 &  53.5 &   9.4 &   49.0 \\ 
 -2.03 &  -5.74 &  3.20 &  0.90 &  46.3 &   8.1 &   64.4 \\ 
 -1.73 &  -5.74 &  2.86 &  0.90 &  50.3 &   9.0 &   67.2 \\ 
 -0.84 &  -5.73 &  2.26 &  0.18 &  55.0 &   2.2 &  237.5 \\ 
 -0.55 &  -5.73 &  1.73 &  0.12 &  52.5 &   2.0 &  223.1 \\ 
 -0.25 &  -5.73 &  2.08 &  0.10 &  47.8 &   1.3 &  237.2 \\ 
  0.05 &  -5.73 &  1.82 &  0.08 &  45.1 &   1.3 &  290.4 \\ 
  0.34 &  -5.73 &  1.60 &  0.11 &  46.1 &   1.9 &  234.0 \\ 
  0.64 &  -5.73 &  2.12 &  0.20 &  32.2 &   2.7 &  157.1 \\ 
  0.94 &  -5.73 &  2.92 &  0.40 &  25.1 &   4.0 &  109.9 \\ 
  1.23 &  -5.73 &  2.27 &  0.84 &  31.2 &  10.9 &   79.8 \\ 
 -2.32 &  -5.44 &  4.59 &  2.23 &  58.1 &  13.8 &   43.8 \\ 
 -2.03 &  -5.44 &  5.52 &  1.91 &  40.6 &   9.5 &   58.9 \\ 
 -0.55 &  -5.43 &  1.92 &  0.28 &  56.6 &   4.0 &  168.6 \\ 
 -0.25 &  -5.43 &  1.54 &  0.20 &  47.9 &   3.7 &  169.9 \\ 
  0.05 &  -5.43 &  1.55 &  0.17 &  49.7 &   3.2 &  190.4 \\ 
  0.34 &  -5.43 &  1.39 &  0.16 &  54.6 &   3.2 &  200.9 \\ 
  0.64 &  -5.43 &  1.34 &  0.22 &  40.8 &   4.7 &  156.4 \\ 
  0.94 &  -5.43 &  1.45 &  0.37 &  24.9 &   7.3 &  101.9 \\ 
  1.23 &  -5.43 &  1.73 &  0.75 &  19.2 &  12.4 &   80.9 \\ 
 -0.55 &  -5.14 &  2.79 &  0.59 &  55.4 &   6.0 &  123.3 \\ 
 -0.25 &  -5.14 &  2.70 &  0.35 &  62.4 &   3.6 &  117.1 \\ 
  0.05 &  -5.14 &  1.57 &  0.30 &  64.3 &   5.5 &  125.3 \\ 
  0.34 &  -5.14 &  1.59 &  0.23 &  62.8 &   4.2 &  146.0 \\ 
  0.64 &  -5.14 &  1.46 &  0.21 &  49.4 &   4.2 &  156.9 \\ 
  1.23 &  -5.14 &  2.43 &  1.07 &  57.2 &  12.6 &   74.9 \\ 
 -0.25 &  -4.84 &  1.93 &  0.92 &  63.6 &  13.6 &   79.3 \\ 
  0.34 &  -4.84 &  1.19 &  0.56 &  65.7 &  13.4 &   97.9 \\ 
  0.64 &  -4.84 &  1.30 &  0.38 &  61.7 &   8.3 &  122.4 \\ 
 \enddata

\tablenotetext{a} {Offsets in arcminutes from $ 5^{\mathrm{h}}32^{\mathrm{m}}50^{\mathrm{s}}$, $ -5\arcdeg15\arcmin00\arcsec$ (1950). Sorted in Dec first.}
 \tablenotetext{b} {Position angle of E-vector in degrees east from north.}
 \tablenotetext{c} {Jy/20$\arcsec$ beam.}

 \end{deluxetable}

%%% Local Variables: 
%%% mode: latex
%%% TeX-master: "omc2"
%%% End: 

 \begin{deluxetable}{rrrrrrr}
 \tabletypesize{\footnotesize}
 \tablecaption{OMC-2, 350 \micron\ Results\label{ta:omc2}}
 \tablecolumns{7}
 \tablewidth{0pt}
 \tablehead{
 \colhead{$\Delta\alpha$ \tablenotemark{a}} &
 \colhead{$\Delta\delta$ \tablenotemark{a}} &
 \colhead{$P$} & \colhead{$\sigma_{P}$} &
 \colhead{$PA$ \tablenotemark{b}} & \colhead{$\sigma_{PA}$} &
 \colhead{Flux \tablenotemark{c}}}

 \startdata
  1.43 &  -0.08 &  1.67 &  0.66 & 105.1 &  10.9 &   24.1 \\ 
  0.83 &   0.22 &  1.25 &  0.62 & 109.0 &  14.0 &   34.9 \\ 
  1.13 &   0.22 &  2.02 &  0.54 & 109.1 &   7.6 &   39.8 \\ 
  1.72 &   0.22 &  2.48 &  0.68 & 125.5 &   7.7 &   28.2 \\ 
  2.02 &   0.22 &  6.47 &  2.90 & 127.8 &  12.7 &   16.4 \\ 
  2.02 &   0.52 &  2.54 &  0.93 & 129.9 &  10.4 &   20.6 \\ 
  1.43 &   0.52 &  0.79 &  0.34 &  93.6 &  12.2 &   48.0 \\ 
  0.83 &   0.81 &  4.64 &  1.06 &  85.0 &   6.6 &   25.9 \\ 
  1.13 &   0.81 &  1.35 &  0.45 & 103.3 &   9.3 &   44.2 \\ 
  1.72 &   0.81 &  1.04 &  0.43 & 130.7 &  11.8 &   50.9 \\ 
  2.32 &   0.81 &  3.56 &  1.34 & 141.4 &  10.4 &   17.8 \\ 
  0.83 &   1.11 &  1.96 &  0.92 & 103.1 &  12.7 &   24.5 \\ 
  1.43 &   1.11 &  1.46 &  0.45 & 137.8 &   8.8 &   51.4 \\ 
  2.02 &   1.11 &  1.70 &  0.72 & 128.8 &  12.2 &   32.5 \\ 
  2.32 &   1.11 &  2.70 &  1.19 & 114.0 &  12.5 &   23.6 \\ 
  1.43 &   1.41 &  2.98 &  0.70 &  80.8 &   6.9 &   40.0 \\ 
  1.72 &   1.41 &  1.58 &  0.67 & 114.4 &  12.0 &   37.0 \\ 
  2.02 &   1.41 &  3.26 &  0.88 & 125.9 &   7.7 &   28.0 \\ 
  1.58 &   1.93 &  3.49 &  1.28 & 123.7 &  10.4 &   43.5 \\ 
  1.58 &   2.52 &  0.95 &  0.42 & 133.0 &  12.8 &   55.9 \\ 
  1.88 &   2.52 &  0.39 &  0.28 & 126.5 &  20.7 &   75.5 \\ 
  2.17 &   2.52 &  0.75 &  0.31 & 111.8 &  12.2 &   89.8 \\ 
  2.47 &   2.52 &  0.62 &  0.25 & 137.2 &  11.4 &   81.2 \\ 
  2.77 &   2.52 &  0.30 &  0.33 &  25.2 &  32.0 &   65.2 \\ 
  1.28 &   2.82 &  0.94 &  0.47 & 167.1 &  14.3 &   62.6 \\ 
  1.58 &   2.82 &  0.05 &  0.27 & 118.3 & 148.5 &   71.2 \\ 
  1.88 &   2.82 &  1.05 &  0.23 & 120.4 &   6.9 &   90.4 \\ 
  2.17 &   2.82 &  0.52 &  0.14 & 104.5 &   7.3 &  143.5 \\ 
  2.47 &   2.82 &  0.25 &  0.15 & 115.6 &  16.7 &  111.4 \\ 
  2.77 &   2.82 &  0.67 &  0.26 & 158.0 &  11.1 &   73.0 \\ 
  1.88 &   3.12 &  0.28 &  0.20 & 147.7 &  22.9 &   99.7 \\ 
  2.17 &   3.12 &  0.14 &  0.08 & 101.3 &  16.7 &  200.0 \\ 
  1.28 &   3.12 &  0.09 &  0.40 & 139.8 & 122.0 &   60.1 \\ 
  1.58 &   3.12 &  0.27 &  0.33 & 165.7 &  36.1 &   68.8 \\ 
  2.47 &   3.12 &  0.36 &  0.13 & 128.1 &  10.0 &  148.9 \\ 
  2.77 &   3.12 &  0.36 &  0.24 & 126.4 &  21.0 &   78.9 \\ 
  1.58 &   3.41 &  1.06 &  0.30 & 172.4 &   8.0 &   70.1 \\ 
  1.88 &   3.41 &  0.70 &  0.26 & 161.7 &  10.5 &   74.8 \\ 
  2.17 &   3.41 &  0.15 &  0.22 &  13.3 &  34.5 &  117.5 \\ 
  2.47 &   3.41 &  0.08 &  0.15 &  88.7 &  50.6 &  136.8 \\ 
  2.77 &   3.41 &  0.39 &  0.29 &  17.9 &  21.1 &   76.3 \\ 
  1.88 &   3.71 &  0.88 &  0.36 &   6.3 &  11.7 &   72.2 \\ 
  2.17 &   3.71 &  0.87 &  0.27 & 159.6 &   8.7 &   99.8 \\ 
  2.47 &   3.71 &  0.46 &  0.23 & 147.7 &  16.0 &   98.6 \\ 
  1.28 &   4.01 &  2.59 &  0.93 &  16.3 &  10.2 &   58.0 \\ 
  1.58 &   4.01 &  1.66 &  0.47 &   4.3 &   8.1 &   65.7 \\ 
  2.17 &   4.01 &  1.15 &  0.45 & 164.7 &  11.0 &   68.2 \\ 
  2.47 &   4.01 &  1.58 &  0.43 & 159.7 &   7.8 &   67.8 \\ 
  1.58 &   4.30 &  1.64 &  0.64 &  26.3 &  11.1 &   79.8 \\ 
  1.88 &   4.30 &  1.60 &  0.66 & 176.9 &  11.6 &   69.4 \\ 
  2.17 &   4.30 &  1.74 &  0.73 & 175.4 &  11.8 &   61.2 \\ 
  2.47 &   4.30 &  2.04 &  0.99 & 146.4 &  13.1 &   54.3 \\ 
 \enddata

\tablenotetext{a} {Offsets in arcminutes from $ 5^{\mathrm{h}}32^{\mathrm{m}}50^{\mathrm{s}}$, $ -5\arcdeg15\arcmin00\arcsec$ (1950). Sorted in Dec first.}
 \tablenotetext{b} {Position angle of E-vector in degrees east from north.}
 \tablenotetext{c} {Jy/20$\arcsec$ beam.}

 \end{deluxetable}

%%% Local Variables: 
%%% mode: latex
%%% TeX-master: "omc3"
%%% End: 

 \begin{deluxetable}{rrrrrrr}
 \tabletypesize{\footnotesize}
 \tablecaption{OMC-3 \& IRAS 05327-457, 350 \micron\ Results\label{ta:omc3}}
 \tablecolumns{7}
 \tablewidth{0pt}
 \tablehead{
 \colhead{$\Delta\alpha$ \tablenotemark{a}} &
 \colhead{$\Delta\delta$ \tablenotemark{a}} &
 \colhead{$P$} & \colhead{$\sigma_{P}$} &
 \colhead{$PA$ \tablenotemark{b}} & \colhead{$\sigma_{PA}$} &
 \colhead{Flux \tablenotemark{c}}}

 \startdata
  1.75 &   6.82 &  1.05 &  0.37 &  96.6 &  10.2 &   23.9 \\ 
  2.05 &   6.82 &  0.60 &  0.24 &  69.8 &  11.5 &   25.1 \\ 
  2.35 &   6.82 &  0.81 &  0.34 &  90.8 &  12.1 &   24.3 \\ 
  1.46 &   7.12 &  1.01 &  0.43 &  85.9 &  12.3 &   21.3 \\ 
  2.35 &   7.12 &  0.72 &  0.23 & 123.0 &   9.2 &   32.0 \\ 
  2.64 &   7.12 &  1.34 &  0.65 & 126.0 &  13.6 &   22.1 \\ 
  2.94 &   7.12 &  1.19 &  0.41 & 119.5 &   9.9 &   18.5 \\ 
  2.35 &   7.42 &  1.42 &  0.34 & 120.5 &   7.0 &   30.0 \\ 
  2.64 &   7.42 &  1.39 &  0.35 & 116.6 &   7.2 &   19.6 \\ 
  2.94 &   7.42 &  1.50 &  0.43 & 124.0 &   8.2 &   16.9 \\ 
  2.05 &   7.42 &  0.22 &  0.17 & 120.9 &  22.3 &   40.0 \\ 
  1.46 &   7.42 &  1.59 &  0.50 & 114.5 &   9.1 &   22.2 \\ 
  1.46 &   7.71 &  2.35 &  0.37 & 105.4 &   4.2 &   26.3 \\ 
  1.75 &   7.71 &  1.20 &  0.26 & 122.0 &   6.2 &   29.6 \\ 
  2.05 &   7.71 &  1.19 &  0.35 & 130.9 &   8.4 &   38.9 \\ 
  2.35 &   7.71 &  1.06 &  0.24 & 133.5 &   6.6 &   37.0 \\ 
  2.64 &   7.71 &  1.95 &  0.36 & 152.5 &   5.4 &   22.6 \\ 
  2.94 &   7.71 &  1.67 &  0.45 & 141.0 &   7.8 &   17.3 \\ 
  1.46 &   8.01 &  1.69 &  0.41 & 116.6 &   6.8 &   20.9 \\ 
  1.75 &   8.01 &  1.68 &  0.59 &  98.6 &   9.9 &   22.5 \\ 
  2.05 &   8.01 &  0.90 &  0.26 & 137.1 &   8.5 &   35.4 \\ 
  2.64 &   8.01 &  1.44 &  0.36 & 126.9 &   7.1 &   23.6 \\ 
  2.94 &   8.01 &  1.96 &  0.67 & 129.8 &   9.7 &   15.6 \\ 
  2.05 &   8.31 &  1.77 &  0.45 & 106.4 &   6.9 &   22.9 \\ 
  2.35 &   8.31 &  1.68 &  0.36 & 131.2 &   6.2 &   25.8 \\ 
  2.64 &   8.31 &  2.13 &  0.45 & 117.6 &   6.0 &   18.1 \\ 
  1.40 &  10.70 &  1.74 &  0.63 & 154.8 &  10.4 &   32.4 \\ 
  1.70 &  10.70 &  1.98 &  0.54 & 148.4 &   7.7 &   43.1 \\ 
  2.00 &  10.70 &  1.76 &  0.63 & 159.5 &  10.3 &   37.5 \\ 
  0.81 &  10.99 &  2.19 &  0.68 & 150.2 &   8.9 &   36.4 \\ 
  1.11 &  10.99 &  2.27 &  0.76 & 131.7 &   9.6 &   34.6 \\ 
  1.40 &  10.99 &  1.77 &  0.58 & 144.3 &   9.4 &   41.2 \\ 
  1.70 &  10.99 &  1.33 &  0.44 & 136.1 &   9.5 &   45.9 \\ 
  0.81 &  11.29 &  2.12 &  0.54 & 150.0 &   7.3 &   44.5 \\ 
  1.11 &  11.29 &  1.88 &  0.56 & 138.2 &   8.4 &   42.4 \\ 
  1.40 &  11.29 &  0.88 &  0.30 & 143.4 &   9.0 &   60.6 \\ 
  2.00 &  11.29 &  1.37 &  0.58 & 143.7 &  12.1 &   37.8 \\ 
  0.22 &  11.59 &  1.78 &  0.84 & 111.2 &  13.6 &   39.7 \\ 
  0.81 &  11.59 &  1.87 &  0.41 & 126.1 &   7.2 &   52.1 \\ 
  1.11 &  11.59 &  1.60 &  0.22 & 132.0 &   4.1 &   68.2 \\ 
  1.70 &  11.59 &  0.74 &  0.34 & 147.0 &  13.3 &   61.7 \\ 
  1.40 &  11.59 &  1.25 &  0.20 & 127.2 &   4.6 &  110.0 \\ 
  2.29 &  11.59 &  2.44 &  1.03 & 168.3 &  12.1 &   31.9 \\ 
  0.22 &  11.88 &  1.37 &  0.56 & 130.5 &  11.9 &   51.2 \\ 
  0.51 &  11.88 &  1.14 &  0.48 & 132.4 &  12.0 &   62.8 \\ 
  0.81 &  11.88 &  1.77 &  0.25 & 137.3 &   4.0 &   67.8 \\ 
  1.11 &  11.88 &  1.54 &  0.23 & 144.2 &   4.2 &   83.8 \\ 
  1.40 &  11.88 &  1.56 &  0.29 & 147.3 &   5.9 &   69.5 \\ 
  1.70 &  11.88 &  1.42 &  0.48 & 146.9 &   9.7 &   47.7 \\ 
 -0.08 &  12.18 &  1.39 &  0.62 & 111.5 &  12.8 &   61.8 \\ 
  0.22 &  12.18 &  1.66 &  0.31 & 134.3 &   5.3 &   89.3 \\ 
  0.51 &  12.18 &  1.43 &  0.26 & 127.9 &   5.4 &   99.4 \\ 
  0.81 &  12.18 &  1.57 &  0.28 & 130.4 &   5.2 &   93.0 \\ 
  1.11 &  12.18 &  1.28 &  0.39 & 139.2 &   9.4 &   71.0 \\ 
  1.40 &  12.18 &  1.28 &  0.46 & 123.7 &  11.9 &   47.3 \\ 
  1.70 &  12.18 &  1.49 &  0.66 & 145.1 &  12.5 &   35.8 \\ 
 -0.08 &  12.48 &  2.32 &  0.57 & 141.4 &   7.0 &   82.2 \\ 
  0.22 &  12.48 &  2.01 &  0.34 & 128.5 &   4.8 &   90.0 \\ 
  0.51 &  12.48 &  1.92 &  0.37 & 126.1 &   5.7 &   74.1 \\ 
  0.81 &  12.48 &  1.25 &  0.47 & 132.7 &  10.8 &   72.0 \\ 
  1.11 &  12.48 &  1.88 &  0.63 & 135.1 &   9.5 &   53.7 \\ 
  1.40 &  12.48 &  2.66 &  0.91 & 145.4 &   9.8 &   40.4 \\ 
 -0.08 &  12.77 &  1.94 &  0.85 & 132.1 &  12.6 &   89.1 \\ 
  0.22 &  12.77 &  2.11 &  0.59 & 136.6 &   8.0 &   66.8 \\ 
  0.51 &  12.77 &  2.41 &  0.69 & 134.0 &   8.2 &   47.5 \\ 
  0.81 &  12.77 &  2.22 &  0.75 & 131.6 &   9.7 &   37.9 \\ 
  1.11 &  12.77 &  2.68 &  0.99 & 140.1 &  10.5 &   28.2 \\ 
  0.22 &  13.07 &  4.11 &  1.40 & 137.8 &   9.8 &   47.2 \\ 
  0.51 &  13.07 &  3.19 &  1.48 & 140.8 &  13.0 &   38.8 \\ 
  0.81 &  13.07 &  4.03 &  1.67 & 131.9 &  12.3 &   31.8 \\ 
 -0.90 &  16.48 &  4.73 &  1.57 & 153.2 &  10.5 &   17.5 \\ 
 -1.20 &  16.77 &  4.50 &  1.62 & 164.2 &  10.6 &   20.8 \\ 
 -0.90 &  16.77 &  3.03 &  0.89 & 147.8 &   8.0 &   21.9 \\ 
 -0.01 &  16.77 &  3.19 &  1.32 & 128.0 &  11.9 &   18.9 \\ 
 -1.49 &  17.07 &  2.52 &  0.97 & 179.1 &  11.1 &   25.2 \\ 
 -0.60 &  17.07 &  3.66 &  0.97 & 125.2 &   7.6 &   19.2 \\ 
 -0.31 &  17.07 &  4.73 &  1.44 & 130.0 &   8.7 &   18.0 \\ 
 -0.01 &  17.07 &  4.16 &  1.19 & 132.9 &   8.2 &   15.8 \\ 
 -0.01 &  17.37 &  4.96 &  1.55 & 150.9 &   8.9 &   15.7 \\ 
 -0.90 &  17.37 &  2.25 &  0.91 & 121.1 &  11.8 &   24.0 \\ 
 -0.31 &  17.37 &  4.23 &  1.32 & 115.9 &   8.9 &   18.1 \\ 
 -1.20 &  17.66 &  2.36 &  0.95 & 157.7 &  11.7 &   23.3 \\ 
 -0.90 &  17.66 &  1.91 &  0.94 & 132.8 &  14.1 &   22.1 \\ 
 -0.60 &  17.66 &  3.40 &  1.14 & 147.3 &   9.0 &   18.5 \\ 
 -0.01 &  17.66 &  3.95 &  1.42 & 147.0 &  10.4 &   14.4 \\ 
 \enddata

\tablenotetext{a} {Offsets in arcminutes from $ 5^{\mathrm{h}}32^{\mathrm{m}}50^{\mathrm{s}}$, $ -5\arcdeg15\arcmin00\arcsec$ (1950). Sorted in Dec first.}
 \tablenotetext{b} {Position angle of E-vector in degrees east from north.}
 \tablenotetext{c} {Jy/20$\arcsec$ beam.}

 \end{deluxetable}

%%% Local Variables: 
%%% mode: latex
%%% TeX-master: "omc4"
%%% End: 

 \begin{deluxetable}{rrrrrrr}
 \tabletypesize{\footnotesize}
 \tablecaption{OMC-4, 350 \micron\ Results\label{ta:omc4}}
 \tablecolumns{7}
 \tablewidth{0pt}
 \tablehead{
 \colhead{$\Delta\alpha$ \tablenotemark{a}} &
 \colhead{$\Delta\delta$ \tablenotemark{a}} &
 \colhead{P} & \colhead{$\sigma_{\rm{P}}$} &
 \colhead{PA \tablenotemark{b}} & \colhead{$\sigma_{\rm{PA}}$} &
 \colhead{Flux \tablenotemark{c}}}

 \startdata
 -3.14 & -23.43 &  1.47 &  0.68 & 160.3 &  13.2 &   25.6 \\ 
 -2.84 & -23.43 &  1.83 &  0.48 & 171.0 &   7.4 &   24.6 \\ 
 -2.55 & -23.43 &  1.96 &  0.68 & 176.7 &   9.7 &   22.7 \\ 
 -3.44 & -23.13 &  1.61 &  0.53 &   0.4 &   9.4 &   33.1 \\ 
 -2.84 & -23.13 &  1.31 &  0.64 & 177.2 &  14.1 &   26.2 \\ 
 -2.55 & -23.13 &  1.83 &  0.47 & 161.3 &   7.3 &   27.5 \\ 
 -3.14 & -22.83 &  1.30 &  0.65 & 166.8 &  14.4 &   29.7 \\ 
 -2.84 & -22.83 &  1.97 &  0.46 & 179.4 &   6.7 &   27.5 \\ 
 -2.55 & -22.83 &  1.90 &  0.65 & 175.6 &   9.8 &   30.9 \\ 
 -2.25 & -22.83 &  0.88 &  0.36 & 178.4 &  12.1 &   35.0 \\ 
 -2.55 & -22.54 &  1.28 &  0.44 & 163.7 &   9.8 &   33.2 \\ 
 -2.25 & -22.54 &  1.34 &  0.39 &   5.7 &   8.5 &   36.9 \\ 
 -1.95 & -22.54 &  2.08 &  0.74 & 165.0 &   8.5 &   26.5 \\ 
 -3.44 & -22.24 &  2.46 &  0.53 & 149.7 &   6.1 &   31.3 \\ 
 -3.14 & -22.24 &  2.00 &  0.76 & 148.9 &  10.8 &   35.0 \\ 
 -2.84 & -22.24 &  1.80 &  0.52 & 158.0 &   8.2 &   31.3 \\ 
 -2.55 & -22.24 &  1.57 &  0.53 & 158.9 &   9.7 &   31.9 \\ 
 -2.25 & -22.24 &  1.81 &  0.41 & 146.3 &   6.5 &   35.3 \\ 
 -2.25 & -21.94 &  1.12 &  0.46 & 145.2 &  11.7 &   34.3 \\ 
 \enddata

\tablenotetext{a} {Offsets in arcminutes from $ 5^{\mathrm{h}}32^{\mathrm{m}}50^{\mathrm{s}}$, $ -5\arcdeg15\arcmin00\arcsec$ (1950). Sorted in Dec first.}
 \tablenotetext{b} {Position angle of E-vector in degrees east from north.}
 \tablenotetext{c} {Jy/20$\arcsec$ beam}

 \end{deluxetable}

\end{document}